\documentclass{aa}
\newcommand {\dr}{$\degr$}
\usepackage{graphicx}
\begin{document}


   \title{Recovery of the global magnetic field configuration of\\
   78 Virginis from Stokes $IQUV$ line profiles}
   \author{V.R. Khalack \inst{1, 2} and G.A. Wade \inst{3}}

   \offprints{V. Khalack \\ \email{khalakv@umoncton.ca}}

   \institute{ D\'{e}partement de physique et d'astronomie, Universit\'{e}
               de Moncton, Moncton, N.-B., E1A 3E9, Canada
               \and
               Main Astronomical Observatory, 27 Zabolotnoho Str., 03680, Kyiv, Ukraine
               \and
               Department of Physics, Royal Military College of Canada, PO Box 17000 stn
               'FORCES', Kingston, Ontario, Canada K7K 4B4
             }

   \date{Received {\it date will be inserted by the editor}\
         Accepted {\it date will be inserted by the editor}
        }

\abstract{The surface magnetic field configuration of the Ap star
HD~118022 (78 Vir) has been reconstructed in the framework of the
magnetic charge distribution (MCD) method from the analysis of
Stokes $IQUV$ spectra obtained using the MuSiCoS
spectropolarimeter at Pic du Midi Observatory.
Magnetically-sensitive Fe~{\sc ii} lines were primarily employed
in the analysis, supposing that iron is evenly distributed over
the stellar surface. We show that the Stokes $IQUV$ profile shapes
and variations of 78 Vir can be approximately fit assuming a
global magnetic field configuration described by a slightly
decentered, inclined magnetic dipole of polar surface intensity
approximately 3.3~kG. The derived inclinations of the stellar
rotational axis to the line of sight $i=24\pm 5\degr$ as well as
to the magnetic dipole axis $\beta=124\pm5\degr$ are in good
agreement with previous estimations by other authors, whereas the
sky-projected position angle\thanks{ $\Omega$ increases clockwise
from the axis to the North Celestial Pole and relates to the
azimuth angle $\Theta$ specified by Landolfi at al.~(\cite{Landolfi+93})
as $\Omega=360\degr-\Theta$.} of the stellar rotation axis
$\Omega\sim110\degr$ is reported here for the first time.
In addition, several lines of Cr~{\sc ii} and Ti~{\sc ii} were
studied, yielding evidence for non-uniform surface distributions
of these elements, and magnetic field results similar to those
derived from Fe.
}

\titlerunning{Recovery of the 78 Vir global magnetic field}

\authorrunning{V.R. Khalack and G.A. Wade}

\maketitle

\section{Introduction
\label{intro}}

78 Virginis (HD~118022) is a bright Ap star, and the first star other than the
sun in which magnetic field was discovered (Babcock~\cite{Babcock47}). The longitudinal
magnetic field of 78 Vir, as first observed by Babcock, is variable, and
constantly negative. 78 Vir is also marginally variable in broadband light,
in radial velocity, and shows variations of its line profiles
(Preston~\cite{Preston69}). As suggested by Preston (\cite{Preston69}), the observed
properties of 78 Vir can be explained in the framework of the oblique rotator
model (Stibbs~\cite{Stibbs50}), in which the star rotates with a period of
approximately 3.7 days and has a rotation axis forming a small angle with
respect to the line of sight.

The first comprehensive modelling of the surface magnetic field structure of
78 Vir was performed by Borra (\cite{Borra80}). Borra obtained high-resolution
measurements of circular polarisation across the Fe\,{\sc ii} $\lambda$4520.2\AA\,
line, which he attempted to reproduce using nine different magnetic field
configuration models. As argued by Borra (\cite{Borra80}), the profiles were
well reproduced with an inclined dipole geometry that contains a moderate
quadrupolar component. Nevertheless, he also pointed out that a decentered
($a$=0.2) dipole model provides {essentially} the same fit quality.
Borra also concluded that the 78 Vir
presents to an observer a remarkably uniform magnetic field over most of its
visible disk, at most phases. It is unfortunate that due to the small
inclination of the rotational axis (about $25\degr$) a significant part of the stellar surface
remains hidden - a part which might have a less uniform field geometry.

The analysis of broadband linear polarization (BBLP) measurements of several Ap
stars (including 78 Vir) by Leroy~et~al.~(\cite{Leroy+96}) suggests that the
magnetic fields of many Ap stars
exhibit departures from the standard oblique rotator model assuming a pure
dipole field geometry. According to those
authors, significant discrepancies between the BBLP observations and the
``canonical model'' results {(Landolfi~et~al. \cite{Landolfi+93})} required the
assumption of local departures from a dipolar field. In particular,
Leroy~et~al. (\cite{Leroy+96}) showed that the BBLP variations of most stars
could be reproduced assuming a dipole magnetic field geometry with slightly
expanded field lines over some parts of the magnetic equator. For 78 Vir, the
observed BBLP variations do not show especially strong departures from the
dipolar case, although according to Leroy~et~al.~(\cite{Leroy+96}) a modified
dipolar model provides a somewhat better fit to the BBLP curves for this star.

In order to better constrain the magnetic field configuration of
78 Vir, in this paper we undertake a more detailed modelling of
the surface magnetic field structure based on the analysis of high
resolution line profiles in all 4 Stokes parameters. In
Sec.~\ref{obs} we discuss the properties of obtained Stokes $IQUV$
spectra, and in Sec.~\ref{fund} we summarise the fundamental
characteristics of 78 Vir. In Sec.~\ref{mod} we describe the
features of the magnetic field modelling framework, and derive the
global magnetic field configuration of the star, comparing
observed and computed Stokes profiles for various spectral
features. In Sec.~\ref{discuss} we discuss the derived global
magnetic field characteristics and their agreement with earlier
studies.

\section{Observations}
\label{obs}

Spectropolarimetric observations of 78 Vir were obtained in 1997
February, 1998 February and 1999 January using the 2m T\'elescope Bernard Lyot
at Observatoire du Pic du Midi (Wade~et~al.~\cite{Wade+00a}). The MuSiCoS
cross-dispersed \'echelle spectrograph (Baudrand \& B\"{o}hm~\cite{BB92}) and
dedicated polarimeter module (Donati et al.~\cite{Donati+99}) were employed for
the observations.

The MuSiCoS spectrograph is a table-top instrument, which allows the
acquisition of a stellar spectrum in a given polarization state (Stokes $V$,
$Q$ or $U$) throughout the spectral range from 4500 to 6600~\AA\ with a resolving
power of about 35 000, in a single exposure. The spectrograph is fed by a
double optical fibre directly from the Cassegrain-mounted polarimeter. The
optical characteristics of the polarization analyser, as well as the
spectropolarimeter observing procedures, are described in detail by Donati et
al.~(\cite{Donati+99}). Observing details specific to the acquisition,
reduction and analysis of the 78 Vir spectra are provided by
Wade~et~al.~(\cite{Wade+00a}).

The journal of spectropolarimetric observations is reported by Wade~et~al.~(\cite{Wade+00a}).
The 52 Stokes $V$, $Q$ and $U$ spectra, obtained on 18 different nights, cover the whole
rotational period of 78 Vir approximately uniformly, and provide an average S/N of about 370.

\section{Fundamental parameters of 78 Vir}

\label{fund}

78 Vir was classified by Cowley et al.~(\cite{C2J2}) as A1pCrSrEu.
Adelman~(\cite{Adelman73a, Adelman73b}) performed both a line
identification and abundance analysis of this star.
The distance to 78 Vir $d=48\pm15$ pcs and its radius $R=1.77\pm0.68R_{\rm \sun}$
are derived by Monier~(\cite{Monier92}) using the Infrared Flux Method. On the
other hand, taking into account the mean angular diameter
$\theta_{\rm a}$= 0.343~{milliarcsec} obtained by
Monier~(\cite{Monier92}) for this star and its distance
$d=56.2\pm2.5$~pc derived from the Hipparcos parallax (ESA~\cite{ESA97}),
we find a somewhat larger value for the stellar radius $R=2.06\pm0.17~R_{\rm \sun}$.
Using the Hipparcos visual magnitude and parallax we have found the absolute
visual magnitude of 78 Vir, $M_{\rm v}= 1.18\pm0.10$. Taking into
account the bolometric correction (Flower~\cite{Flower96}), which
corresponds to Monier's (1992) $9200\pm 290$~K effective temperature
and the bolometric zeropoint correction BC$_{\rm V}$=-0.07,
which is obtained assuming $M_{bol}^{\sun}$=4.74 for
the Sun (Bessell~et~al.~\cite{Bessell+98}), we can find
the absolute luminosity for 78 Vir $L_{\rm \star}=(27.3\pm2.5)L_{\rm \sun}$.
For 78 Vir Monier~(\cite{Monier92}) also estimates the integrated flux
$f_{\star}=(2.7\pm 0.27)\times 10^{-7} erg\; sm^{-2} s^{-1}$. Taking into account
the distance to the star, this provides the luminosity
$L_{\rm \star}=(26.5\pm2.8)L_{\rm \sun}$.
Finally, employing the stellar radius and
the effective temperature (see Table~\ref{tab2})
we obtain a third estimate of the absolute luminosity
$L_{\rm \star}=27.4\pm4.5L_{\rm \sun}$. All of these values are in good mutual agreement.
The luminosity $L_{\rm \star}=(27.3\pm2.5)~L_{\rm \sun}$
together with the effective temperature (see Table~\ref{tab2}) allow us to
find the stellar mass $M_{\rm \star}=2.18\pm0.06M_{\rm \sun}$ and age
$3.0^{+0.7}_{-1.3}\times10^{8}$ years (see Fig.~\ref{HRdiaram}) by
spline interpolation (Sandwell~\cite{Sandwell87})
in the model evolutionary tracks of Schaller~et~al.~(\cite{Schaller+92})
for metallicity $Z$=0.02. The derived age suggests that 78 Vir
has completed approximately 37\% of its main sequence life.
The mass and radius provide $\log g = 4.16\pm 0.07$.
This value is marginally inconsistent with that of the Monier~(\cite{Monier92}), but it
is in good agreement with the value $\log g =4.20$ reported by King~et~al.~(\cite{King+03}).
A recent investigation of the Ursa Major stream characteristics performed by
King~et~al.~(\cite{King+03}) derived $Z=0.016-0.02$ and age $(5\pm1)\times10^{8}$ years
for the stream. They found that 78 Vir is a "certain member" of the stream
from photometric data, while the kinematic characteristics of the star argue
for its "probable non-membership".
The derived age for 78 Vir is marginally inconsistent with
the UMa stream age. The published metallicity of 78 Vir
is $\log(Fe/H)_{\star}-\log(Fe/H)_{\sun}=^{-0.13}_{+1.58}$
(Cayrel~de~Strobel~et~al.~\cite{Cayrel+97}) and it is
not clear if this star truly has a solar metallicity. For this reason the
derived uncertainties of the mass and age may be somewhat larger than indicated here.

\begin{table}[t]
\caption[]{Fundamental parameters of 78 Vir.}
\begin{tabular}{lcc}
\hline
Parameter    &   Value      &  Reference\\
\hline
Spectral type& A1pCrSrEu&Cowley et al.~(\cite{C2J2})\\
Age          & $3.0^{+0.7}_{-1.3}\times10^{8}$y.& this work\\
Distance     & 56.2$\pm$2.5 pcs& Hipparcos (ESA~\cite{ESA97})\\
Period       & 3$.^{d}$7220$\pm0.^{d}$003 & Preston~(\cite{Preston69}) \\
$T_{\rm eff}$&9200$\pm$290K  &Monier~(\cite{Monier92}) \\
$\log{g}$    &4.50$\pm$ 0.25 & ibidem \\
$M_{\rm v}$    & 1.18$\pm$0.10         &this work\\
$L_{\rm \star}$& 27.3$\pm$2.5 $L_{\sun}$&this work\\
$R_{\rm \star}$& 2.06$\pm$0.17$R_{\sun}$& this work\\
$M_{\rm \star}$& 2.18$\pm$0.06$M_{\sun}$& this work \\
$V_{\rm e}\sin{i}$& 12$\pm$1 km s$^{-1}$& this work \\
$i$          & 25$\degr\pm$5$\degr$ &Leroy~et~al.~(\cite{Leroy+96}) \\
$\beta$      & 120$\degr\pm$5$\degr$&ibidem \\
\hline
\end{tabular} \label{tab2}
\end{table}


Babcock~(\cite{Babcock47}) observed that the longitudinal magnetic field of
78 Vir was variable, but always negative. Analysing Babcock's data along with
his own measurements, Preston~(\cite{Preston69}) determined
the periodic character of the magnetic field variability and derived the
ephemeris:

\begin{equation}\label{ephemeris}
{\rm JD}\ {\rm (magnetic\;\; maximum)}= 2 434 816.9 + 3.^{d}7220\cdot {\rm E}.
\end{equation}

\noindent Several authors have confirmed this period on the basis of further
magnetic field measurements (Wolff~\&~Wolff~\cite{W+W71};
Wolff~\&~Bonsack~\cite{W+B72}; Wolff~\cite{Wolff78}; Borra~\cite{Borra80};
Borra~\& Landstreet~\cite{B+L80}; Landstreet~\cite{Landstreet82};
{Wade~et~al.~\cite{Wade+00b}), while
Landstreet~(private communication) has estimated the accuracy of the period
determination to be $\pm0.^{d}003$. Preston~(\cite{Preston69}) also found that
the crossover effect is very pronounced and is present throughout most of the
magnetic cycle. This feature is particularly remarkable at phase 0.85
(Wolff~\&~Bonsack~\cite{W+B72}, see also Wade et al.~\cite{Wade+00a}).

Applying the oblique rotator model with the distorted dipole
approximation, 
Leroy~(private communication) found that they must adopt a period 3.7218 days
in order to fit the observed BBLP variations. As discussed in Sect.~\ref{intro},
their model, using local modifications of the axisymmetric magnetic field imposed
near the magnetic equator, differs only mildly from a dipole configuration. The
derived rotational axis inclination and dipole obliquity $i=25\degr$ and
$\beta=120\degr$ provide a good match to the longitudinal field and linear
polarisation variations. Nevertheless, the period inferred is somewhat shorter
than that obtained by other authors (Eq.~\ref{ephemeris}), and
Wade~et~al.~(\cite{Wade+00b}) reported that this period is only barely
consistent with the available longitudinal field measurements.

\begin{figure}[t]
\includegraphics[angle=-90,width=3.5in]{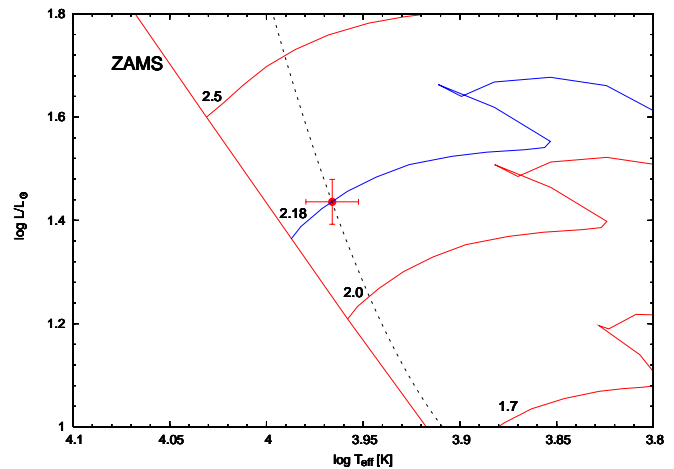}
\caption{The position of 78 Vir on the H-R diagram is compared with
the theoretical evolutionary tracks of Schaller~et~al.~(\cite{Schaller+92}),
computed for solar metallicity and masses M = 1.7, 2.0 and 2.5 $M_{\rm \sun}$.
The age $3.0\times 10^8$ years and the mass $M_{\rm \star}$ = 2.18 $M_{\rm \sun}$
of the star are estimated by interpolation of the theoretical evolutionary tracks.
The dashed line shows the isochrone, which corresponds to the age of 78 Vir ($3.0\times 10^8$~y).}
\label{HRdiaram}
\end{figure}

78 Vir also shows photometric variability in the visible and the infrared
with the same period (Eq.~\ref{ephemeris}). Combining the visual flux
variability in the $uvby$ bands, obtained by Wolff~\&~Wolff~(\cite{W+W71}), with their own
new observations, Catalano~\&~Leone~(\cite{C+L94}) have tried to improve the
period determination. Their analysis results in the new ephemeris

\begin{equation}\label{ephemeris-new}
{\rm JD}\ {\rm (y\; min.)}=2 434 816 + (3.^{d}722084 \pm 0.^{d}000042) {\rm E}, 
\end{equation}

\noindent which was recently confirmed by Leone~\& Catanzaro~(\cite{L+C01}).
These authors have shown with the help of Hipparcos photometry that the light
variations of 78 Vir are not purely sinusoidal. Taking into account all
archival magnetic field observations as well as their own data,
Leone~\&~Catanzaro~(\cite{L+C01}) have shown that the longitudinal field
measurements are in phase only when the 3.722084 day period is adopted.
Moreover, by adopting the 3.7218 day period suggested by
Leroy~(private communication), they find a 0.12-cycle phase shift between the
Hipparcos light curves and the other light curves.

Monier~(\cite{Monier92}) has derived for 78 Vir from the simulation
of the energy distribution in the spectral range from 1200\AA\, to 22000\AA\, an
effective temperature $T_{\rm eff}=9200\pm290$~K, a surface gravity $\log{g}=4.5$
and a photospheric metallicity $[M/H]=10$. In addition, using a model atmosphere
with these values of $\log{g}$ and $[M/H]$, he has found
{that the effective temperature $T_{\rm eff}=9300$K provides the best
description of the spectral energy distribution in the region from 1200\AA\,
to 8000\AA\, at rotational phase 0.0, while $T_{\rm eff}=9200$K seems to
be the best effective temperature, at phase 0.5. These temperatures are
somewhat lower than most previous estimates, ranging from 9700K to 10700K
(Mihalas~\&~Henshaw~\cite{M+H66}; Wolff~\cite{Wolff67};
Jugaku~\&~Sargent~\cite{J+S68}), probably because the models used by these
authors were unblanketed and calculated for solar abundances.

The fundamental parameters of 78 Vir are summarised in Table~\ref{tab2}.


\section{Modelling of the Stokes $IQUV$ spectra}
\label{mod}

According to the discussion of Sect.~\ref{fund}, an ATLAS9 model
(Kurucz~\cite{Kurucz94}) with parameters $T_{\rm eff}=9250$K,
$\log{g}=4.5$, [M/H]=0 and
microturbulent velocity $v_{\rm t}$=0 km s$^{-1}$ 
successfully approximates the stellar atmosphere of 78 Vir.

We describe the structure of the surface magnetic field of 78 Vir (assumed to
be a rigidly rotating and spherically symmetric star) using the {\em magnetic
charge distribution} (MCD) method (Gerth~et~al.~\cite{gerth+};
Khalack~et~al.~\cite{khalack+}). Originally, the MCD method considered a
system of spatially separated {\it point field sources} with {\it virtual
magnetic charges} in the stellar interior. {In order to provide zero magnetic
flux through the stellar surface the sum of ``magnetic charges'' should be kept to zero.}
These sources produce a magnetic
field whose potential at each point on the stellar surface is specified by
the superposition of potentials of individual sources (Gerth~et~al.~\cite{gerth+};
Gerth~\&~Glagolevskij~\cite{gerth+01, gerth+04}). Since the number of sources is usually
more than one, we actually operate with a system of several magnetic dipoles
located in the stellar interior. In order to minimize the number of free model
parameters, the most convenient way is to consider a system of two sources, which
mathematically formulate a magnetic dipole (Khalack~et~al.~\cite{khalack+}).
When the dipole is centered on the stellar centre and the dipole size is
much smaller than the stellar radius (Khalack~\cite{khalack02}), the MCD model
transforms to the conventional model of a symmetric inclined magnetic rotator
(Stibbs~\cite{Stibbs50}). Otherwise, we deal with a more complex magnetic
field configuration.
The mathematical verification of the MCD model as well as the procedure of
specification of the angle $\beta$ between the rotational and magnetic dipole
axes, the coordinates and the field strengths of the positive and negative
magnetic poles on the basis of the derived free model parameters is described
in detail by Khalack~et~al.~(\cite{khalack+03}).

\begin{table}[t]
\caption{Here the individual columns specify
the ion, the wavelength, the quantum number $J$ and the Land\'e factor
for lower and upper atomic levels, the observed Stokes $I$ profile depth and the
line scaling factor (LSF) (see Sect.~\ref{integral}), derived from the
comparison of simulated equivalent width of the Stokes $Q$ and $U$ Zeeman profiles
with the BBLP data (Leroy~\cite{Leroy95}). Asterisks marks the
LSF calculated from Stokes $Q$ and $U$ profiles simulated for a field
structure that is derived from only Stokes $I$ and $V$ profile variability
(see Table.~\ref{fe2sm}).}

\begin{tabular}{lccccccl}

\hline

Ion& $\lambda$, \AA &$J_{\rm lo}$&$g_{\rm lo}$&$J_{\rm up}$&$g_{\rm up}$&$I_{\rm obs}$&LSF\\

\hline

Fe\,{\sc ii} & 4620.52& 3.5 & 1.21 & 3.5 & 1.40 & 0.33&0.035\\


Fe\,{\sc ii} & 4635.32& 2.5 & 1.20 & 3.5 & 1.13 &0.32&0.036$^*$\\

Fe\,{\sc i}  & 4635.85& 1.0 & 1.49 & 2.0 & 1.89 &    &      \\


Fe\,{\sc ii} & 4923.93& 2.5 & 2.00 & 1.5 & 2.40 &0.58&0.079\\


Ti\,{\sc i}  & 5017.95& 4.0 & 1.50 & 5.0 & 1.18 &    & \\

Cr\,{\sc i}  & 5018.15& 2.0 & 1.16 & 3.0 & 1.09 &    & \\

Fe\,{\sc ii} & 5018.44& 2.5 & 2.00 & 2.5 & 1.87 &0.60&0.102\\

Cr\,{\sc ii} & 5018.84& 3.5 & 1.24 & 3.5 & 1.42 &    & \\


Fe\,{\sc ii} & 5100.61& 4.5 & 1.54 & 4.5 & 1.34 &    & \\

Fe\,{\sc ii} & 5100.66& 4.5 & 1.31 & 3.5 & 1.40 &    & \\

Fe\,{\sc ii} & 5100.73& 4.5 & 1.54 & 5.5 & 1.36 &    & \\

Fe\,{\sc ii} & 5100.85& 1.5 & 0.80 & 1.5 & 0.72 &0.47&0.083$^*$\\


Fe\,{\sc i}  & 5168.90& 3.0 & 1.50 & 3.0 & 1.75 &    & \\

Fe\,{\sc ii} & 5169.03& 2.5 & 2.00 & 3.5 & 1.70 &0.59&0.178\\

Fe\,{\sc i}  & 5169.30& 4.0 & 1.26 & 3.0 & 1.32 &    & \\


Fe\,{\sc ii} & 5197.48& 2.5 & 1.20 & 1.5 & 0.72 &    & \\

Fe\,{\sc ii} & 5197.58& 2.5 & 0.57 & 1.5 & 0.44 &0.39&0.073\\


Fe\,{\sc ii} & 5362.74& 3.5 & 1.15 & 4.5 & 1.24 &    & \\

Fe\,{\sc ii} & 5362.87& 4.5 & 1.15 & 3.5 & 1.40 &0.45&0.112$^*$\\

Cr\,{\sc i}  & 5362.96& 3.0 & 1.33 & 3.0 & 1.06 &    & \\

Fe\,{\sc ii} & 5362.97& 3.5 & 1.44 & 4.5 & 1.34 &    &\\

Cr\,{\sc ii} & 5363.88& 3.5 & 1.71 & 3.5 & 1.39 &    & \\


Fe\,{\sc ii} & 6247.35& 1.5 & 0.40 & 2.5 & 0.62 &    & \\

Fe\,{\sc ii} & 6247.56& 2.5 & 1.33 & 1.5 & 1.72 &0.36&0.036$^*$\\


Fe\,{\sc ii} & 6432.68& 2.5 & 2.00 & 2.5 & 1.65 &0.24&0.026\\


Fe\,{\sc ii} & 6516.08& 2.5 & 2.00 & 3.5 & 1.58 &0.25&0.032\\

\hline

\end{tabular} \label{tab3}

\end{table}

\subsection{Procedure
\label{proc}}

From previous analyses of the Stokes $IQUV$ spectra of 78 Vir, it has been
found by Wade~et~al.~(\cite{Wade+00a})
that the {three} strong Fe\,{\sc ii}
lines $\lambda$4923.93\AA, $\lambda$5018.44\AA\, {and $\lambda$5169.03\AA\,}
show the strongest Zeeman signatures of any lines in the optical spectra of
CP stars. The real magnetic field structure of 78 Vir is expected to differ only
marginally from a centered magnetic dipole (based on previous modeling efforts:
Borra~\cite{Borra80}).
The most attention is therefore paid
to the variability of the Stokes $I$ and $V$ profiles, which contain the most information
about the global surface magnetic field configuration.
The Stokes $I$ variation provides information about the surface distribution of
the chemical abundance, as well as the stellar radial velocity $V_{\rm r}$ and
projected rotational velocity $V_{\rm e}\sin{i}$.
The Stokes $V$ profiles are sensitive to the global field configuration,
in particular the lower-order multipolar components. The Stokes $Q$ and $U$ profiles
also provide some constraint on the global field morphology, but are most sensitive to
the smaller-scale structure of the field. Unfortunately, the available
Stokes $Q$ and $U$ profiles have low relative S/N ratio (Wade~et~al.~\cite{Wade+00a}),
and in this study they are used primarily to check the results of the Stokes $I$ and $V$
profile simulations
and to determine the sky-projected position angle of the rotational
axis, $\Omega$. According to Khalack et al.~(\cite{khalack+, khalack+03})
this angle ($0\leq\Omega<360\degr$)
is counted clockwise from the rotation axis to the North
Celestial Pole in the plane of the sky, and is related to the azimuth
angle $\Theta$ specified by Landolfi et al.~(\cite{Landolfi+93}) by
$\Omega=360\degr-\Theta$. The opposite counting of the position angle
in the MCD model is compensated for by the negative sign in
definition of the $B_{\rm y}$-component of the local magnetic field.
Hence a right-handed reference frame is applicable.

We have examined the spectra for additional lines with
prominent polarisation signatures, in order to compile a list of lines
that are especially appropriate for this task. To perform the line
identification we use the VALD-2 resources (Kupka~et~al.~\cite{Kupka+99};
Ryabchikova~et~al.~\cite{Ryab+99}). The adopted list contains Fe\,{\sc ii},
Fe\,{\sc i}, Cr\,{\sc ii}, Cr\,{\sc i}, Ti\,{\sc ii}, Ti\,{\sc i} and
Mg\,{\sc ii} lines, but for the present study we will concentrate primarily on the
strongest Fe\,{\sc ii} lines. The main reason for this is that Fe is presumably
almost uniformly distributed over the stellar surface (see Sect.~\ref{res}) and
in this way we exclude from the fitting procedure a model of the surface
abundance distribution. Table~\ref{tab3} presents the final list of Fe\,{\sc ii}
lines used in the simulation procedure, together with lines of some other
elements that are responsible for blends.

Some of the Fe\,{\sc ii} lines listed in Table~\ref{tab3} have
no detectable features in the Stokes $Q$ and $U$ spectra. Nevertheless, they are
comparatively strong lines (in the Stokes $I$ spectra) with clear
variability in the Stokes $V$ spectra, and are analysed without the linear
polarization data in order to check our final results.

All line profiles are simulated using the {\sc Zeeman2} polarised spectrum
synthesis code (Landstreet~\cite{Landstreet88}; Wade~et~al.~\cite{Wade+01}).
The code has been modified to include a magnetic field described within the
framework of the MCD method (Khalack~et~al.~\cite{khalack+03}), and to allow
for an automatic minimization of the model parameters using the {\it downhill
simplex method} (Press~et~al.~\cite{press+}).
The relatively poor efficiency of the downhill simplex method, requiring a large
number of function evaluations, is a well-known problem.
Repeating the minimization routine 3$\div$4 times in the vicinity of a supposed
minimum in the parameter space allows us to check if the method
converges to a global minimum.
In our case, this technique requires a
comparatively long computational time due to the large amount of analysed
observational data and the large number of free parameters.


\begin{table*}[th]

\parbox[t]{\textwidth}{

\caption[]{Results of Fe\,{\sc ii} lines simulation for $T_{\rm eff}=9250$K,
$\log{g}=4.5$ and $v_{\rm t}$=0 km~s$^{-1}$. 
The first column indicates the free parameters, while the other columns specify
the parameter values for the given line profile. The last
column provides the averaged estimation errors for each parameter.
The first 6 rows show the fit
quality for the each analysed Stokes profile (Eq.~\ref{chi2}), weighted and
unweighted (Eqs.~\ref{chi2wa}-\ref{chi2o}) $\chi^2$-function.
The following 12 rows show the best fit values of the free
model parameters, while the final 9 rows provide the characteristics of the
magnetic dipole poles, which are derived from the free parameters. }

\label{fe2sm}

\vspace{0.in}

\begin{tabular}{l|cccccccccccc}

\hline\hline
Line,  \AA   & 4620 & 4635 & 4923 & 5018 & 5100 & 5169 & 5197 & 5362 & 6247 & 6432 & 6516 &
$\sigma_{\rm er}$\\

\hline

$\chi^2_{I}$ &  2.92& 10.91&  8.96&  6.97&  7.40& 13.10& 16.06&  5.54&  2.27&  8.83&  5.94&  \\

4$\chi^2_{V}$&  6.14&  7.86&  9.14&  8.69&  6.97&  9.09&  8.71&  6.64&  5.00& 12.24&  7.39& \\

6$\chi^2_{Q}$&  7.75&   -  &  7.73&  5.76&   -  &  8.23&  6.35&   -  &   -  & 12.15&  9.30& \\

6$\chi^2_{U}$&  7.42&   -  & 10.04&  6.82&   -  & 10.79&  6.80&   -  &   -  &  9.99&  6.19& \\

$\chi^2_w$   &  6.06&  9.39&  8.97&  7.06&  7.19& 10.30&  9.45&  6.09&  3.64& 10.81&  7.20& \\

$\chi^2$     &  1.75&  6.44&  3.55&  2.81&  4.57&  4.64&  5.11&  3.60&  1.76&  3.90&  2.59& \\

\hline

Qr, kG       &  180 &  254 &  303 &  202 &  230 &  221 &  151 &  192 &  398 &  173 &  289 & 50\\

$a_{1}, 10^{-3}$&8.4&  6.7 &  4.3 &  6.7 &  5.3 &  7.3 & 12.6 &  9.7 &  5.4 &  8.9 &  6.4 & 1.2\\

$\lambda_{1}$& 13\dr& 24\dr& 34\dr& 41\dr& 33\dr& 15\dr& -4\dr& 19\dr& 69\dr& 23\dr& 20\dr& 5\dr\\

$\delta_{1}$ &-45\dr&-46\dr&-52\dr&-51\dr&-60\dr&-30\dr&-40\dr&-44\dr&-39\dr&-45\dr&-39\dr& 8\dr\\

$a_{2}, 10^{-3}$&3.1&  3.7 &  3.5 &  5.2 &  4.3 &  2.5 &  1.6 &  3.7 &  5.1 &  3.8 &  2.7 & 0.3\\

$\lambda_{2}$&118\dr&125\dr&144\dr&135\dr&145\dr& 85\dr&111\dr&101\dr&113\dr& 96\dr&111\dr& 10\dr\\

$\delta_{2}$ &-16\dr& -9\dr& -9\dr&-14\dr&-12\dr&-11\dr&-11\dr&-26\dr&-11\dr&-27\dr&-11\dr& 10\dr\\


$\Omega$     &109\dr&   -  &109\dr&109\dr&   -  &109\dr&102\dr&   -  &   -  &164\dr&128\dr& 17\dr\\

$i$          & 22\dr& 24\dr& 22\dr& 23\dr& 25\dr& 21\dr& 29\dr& 27\dr& 26\dr& 29\dr& 24\dr& 5\dr\\

$\log(Fe/N_{\rm tot})$& -3.36& -2.82& -3.26& -3.24& -3.20& -3.12& -3.25& -2.86& -3.17& -2.97& -3.50& 0.22\\

$V_r$        & -8.78& -8.42& -7.82& -7.51& -8.69& -7.23& -7.93& -8.25& -7.99& -6.57& -6.74& 1.18\\

$V_{\rm e}\sin{i}$& 10.9 & 11.1 & 11.9 & 12.0 & 11.5 &  12.6& 11.9 & 13.0 & 11.9 & 13.7 & 11.5 & 0.9\\

\hline

$\beta$      &125\dr&124\dr&122\dr&120\dr&123\dr&118\dr&129\dr&126\dr&123\dr&126\dr&122\dr& 5\dr\\

$a_{0}, 10^{-3}$&4.5&  3.9 &  2.9 &  6.0 &  3.4 &  4.3 &  6.5 &  5.8 &  4.8 &  5.6 &  3.6 & 1.6 \\

$a, 10^{-3}$ &  4.5 &  3.8 &  2.6 &  0.7 &  3.4 &  3.4 &  6.2 &  4.4 &  2.1 &  3.8 &  3.4 & 1.6\\

$B_{p}$, kG  & 3.25 & 3.95 & 3.47 & 3.18 & 3.13 & 3.02 & 4.00 & 3.45 & 3.42 & 2.69 & 3.92 & 0.7\\

$\lambda_{p}$& -9\dr&-10\dr& -6\dr& -7\dr&-10\dr& -8\dr&-11\dr& -8\dr&-11\dr& -8\dr& -8\dr& 5\dr\\

$\delta_{p}$ &-35\dr&-33\dr&-30\dr&-30\dr&-33\dr&-28\dr&-37\dr&-36\dr&-34\dr&-36\dr&-32\dr& 12\dr\\

$B_{n}$, kG  & -3.18& -3.82& -3.46& -3.16& -3.12& -2.95& -3.85& -3.36& -3.41& -2.62& -3.86& 0.7\\

$\lambda_{n}$&170\dr&170\dr&174\dr&173\dr&170\dr&172\dr&168\dr&171\dr&169\dr&171\dr&172\dr& 5\dr\\

$\delta_{n}$ & 35\dr& 33\dr& 30\dr& 30\dr& 33\dr& 28\dr& 37\dr& 35\dr& 33\dr& 36\dr& 32\dr& 12\dr\\

\hline\hline

\end{tabular}

}

\end{table*}

The free model parameters employed in the line profile simulation can be
divided into two groups: those that describe the magnetic field structure, and
those that describe the stellar geometry and atmospheric characteristics. The
first group includes the parameters  $Q_{\rm r}$, $a_{\rm 1}$ and $a_{\rm 2}$
(which specify the modulus of ``magnetic charges" and their distance from the
center of the star, expressed in the units of stellar radius), and the
parameters $\lambda_{\rm 1}, \delta_{\rm 1}$ and $\lambda_{\rm 2}, \delta_{\rm
2}$ (which specify the spherical coordinates of the ``charges" in the rotational
reference frame of the star). Meanwhile, the second group includes the
parameters $\Omega$ (the position angle),
$i$ (the rotational axis inclination with respect to the line
of sight), $V_{\rm r}$ (the heliocentric radial velocity), $V_{\rm e}\sin{i}$
(the projected rotational velocity), and $\log(N_{\rm x}/N_{\rm tot})$ (the
abundance(s) of the element(s) forming the line to be modeled). Instead of
$V_{\rm e}\sin{i}$ we can obviously use only $V_{\rm e}$ as a free parameter,
but $V_{\rm e}\sin{i}$ is preferable here, because it allows us to compare our
estimate of this parameter directly with the results of other authors. The
minimum number of free model parameters is 11 for the MCD model if we do not
take into account the linear polarization data, and 12 if we do. Sometimes, to
obtain a good fit to the line profiles we need to include in the simulation
lines of other chemical elements (see Table~\ref{tab3}). In this case the
number of free model parameters grows with the number of chemical elements
under investigation, but usually does not exceed 15 parameters.

Initially we specify arbitrarily the values of the free model parameters in the
range of their possible values and simulate the Stokes $IQUV$ spectra at the
phases of the observations. We then compare the simulated profiles with the
observed spectra, and calculate the reduced $\chi^\mathrm{2}$, which we adopt
as a measure of the fit quality. The expression for the $\chi^\mathrm{2}$
function, which reflects the agreement between (for example) the simulated
Stokes $I$ ($I_{\rm \lambda_j}(\varphi_{\rm i})$) and observed Stokes $I$
($I^{obs}_{\rm i,\lambda_j}$) spectra is  given by:

\begin{equation} \label{chi2}
\chi^{2}_{I} = \displaystyle \frac{1}{N_{I}} \sum^{N_{I}}_{i=1}
\displaystyle \frac{1}{N_{i}} \sum^{N_{i}}_{j=1}
\left( \displaystyle \frac{I^{obs}_{i,\lambda_j}-I_{\lambda_j}(\varphi_{i})}
{\sigma[I^{obs}_{i,\lambda_j}]}\right)^{2},
\end{equation}

\noindent where $\sigma[I^{obs}_{\rm i,\lambda_j}]$ corresponds to the measurement errors,
$\varphi_{\rm i}$ is the rotational phase of the observation, $N_{\rm I}$
specifies the number of spectra, while $N_{\rm i}$ is the number of pixels in
each analysed line profile.
In fact, although the simulated spectra are calculated with approximately the
same resolving power as the observed spectra, this does not provide a
direct coincidence of wavelengths in the simulated and observed spectra.
Therefore, during the $\chi^\mathrm{2}$ function evaluation the simulated
spectral intensity at the exact observed wavelength is calculated using a
linear interpolation.

The scale of variability and the measurement errors are different for each of
the four Stokes $IQUV$ spectra (Wade~et~al.~\cite{Wade+00a}). In order to
balance (from a statistical point of view) the information flow from each of the
four Stokes spectra, {the respective $\chi^\mathrm{2}$ contributions are weighted.


The weights are derived from the comparison of the best fit values of
$\chi^2_{\rm V}, \chi^2_{\rm Q}, \chi^2_{\rm U}$ with $\chi^2_{\rm I}$
(when the ordinary $\chi^{2}$ function is minimized) and provide approximately
the same weighted value of these functions for the majority of analysed lines
(see Table~\ref{fe2sm}). The weighted $\chi^\mathrm{2}_{\rm w}$
functions is specified} in the following way:

\begin{equation} \label{chi2wa}
\chi^{2}_{w}=\displaystyle \frac{1}{4}[\chi^{2}_{I}+
4\chi^{2}_{V}+6(\chi^{2}_{Q}+\chi^{2}_{U})],
\end{equation}

\noindent or

\begin{equation} \label{chi2wIV}
\chi^{2}_{w}=\displaystyle \frac{1}{2}[\chi^{2}_{I}+4\chi^{2}_{V}],
\end{equation}

\noindent if we work only with the Stokes $I$ and $V$ spectra. The value of the
$\chi^{2}_{\rm w}$ function is reduced throughout the minimization procedure,
varying the values of all free model parameters. When it reaches its
global minimum, we can also calculate the ordinary $\chi^{2}$ function:

\begin{equation} \label{chi2o}
\chi^{2}=\displaystyle \frac{1}{4}[\chi^{2}_{I}+
\chi^{2}_{V}+\chi^{2}_{Q}+\chi^{2}_{U}].
\end{equation}

\noindent We begin by simulating the strong
Fe\,{\sc ii} $\lambda$4923.927~\AA, $\lambda$5018.44
and $\lambda$5169.03 lines, for which the minimization process
has been repeated for 13$\div$15 times starting from different locations
in the free model parameter space in order to avoid local minima.
In the case of 78 Vir, the downhill simplex method converges to
four global minima, which are caused by the decentered dipole model
symmetry in the MCD method. Two global minima correspond to the
parameter sets $i, \beta, \Omega$ and $i, \beta, 180\degr+\Omega$
which differ only by the angle $\Omega$ and provide exactly the same value
of the $\chi^{2}_{\rm w}$-function. Models with the opposite direction of stellar
rotation ($180\degr-i, 180\degr-\beta, \Omega$ or
$180\degr-i, 180\degr-\beta, 180\degr+\Omega$) do not result in minima.
The other configurations ($i, 180\degr-\beta, \Omega$ and $i,
180\degr-\beta, 180\degr+\Omega$) correspond to models in which
the positive magnetic pole faces the observer. These
configurations are not applicable to the case of 78 Vir, where we
always see the negative magnetic pole, while the positive pole is
partially visible just for phases close to $\varphi=0.0$
(Babcock~\cite{Babcock47}; Borra~\cite{Borra80}). The other two
global minima have parameter sets which differ from the previous
ones by the location of the ``magnetic charges". The two
``magnetic charges" can be located in two different stellar
hemispheres (separated by the equatorial plane), or in a single
hemisphere. If they are not significantly shifted from the stellar
centre, they can form the same dipole axis and will produce almost
the same surface magnetic field configuration. The
differences between the two models can be revealed by the
sensitivity of $\chi^{2}_{\rm w}$ function on the parameter
values, but the minima will still exist. Preference is given to
the deepest global minimum, obtained for the model with the two
``magnetic charges" located in the same stellar hemisphere.
For the other lines the minimization process was performed only
3$\div$4 times (although if we tested the contribution of blends
to the analysed profiles, we ran the minimization routine several
more times). Supposing that all analysed lines contain the
signatures of the same magnetic field structure, we chose the
initial locations in parameter space not far from the parameter
set obtained from the strong line analysis.

In order to evaluate the fit errors (and therefore the uncertainties on the
derived free parameters), we calculate deviations of the simulated profiles
produced as a result of small variations of each of the free parameters, thus
introducing a small shift along one axis in the $\chi^\mathrm{2}$ hyper-space
from the point of the function minimum value. Using this procedure, and taking
into account the uncertainties of the observational data and the obtained
minimal value of $\chi^\mathrm{2}$-function, we can estimate the errors of the
best-fit parameters.

\subsection{Results}

\label{res}


Our initial assumption that iron is uniformly distributed over the surface of
78 Vir appears to be valid, given that the analysed Fe\,{\sc ii} lines reveal
no significant variability of the Stokes $I$ profiles with
rotational phase. Each Fe\,{\sc ii} line (or group of lines) shown in
Table~\ref{tab2} was analyzed independently of the others, using a
stellar atmosphere model with
$T_{\rm eff}=9250$K, $\log{g}=4.5$, $v_{\rm t}$=0 km s$^{-1}$.
For each best-fit simulation the derived free model parameters are given in
Table~\ref{fe2sm}.
During the simulation of some Fe\,{\sc ii} lines the
Stokes $Q$ and $U$ profiles were not taken into account,
and consequently the angle $\Omega$ is not determined for those lines.

\begin{figure*}[th]
\parbox[t]{\textwidth}{
\centerline{%
\begin{tabular}{@{\hspace{-0.22in}}c@{\hspace{-0.23in}}c@{\hspace{-0.23in}}c@{\hspace{-0.23in}}c}
a)~$I/I_{\rm c}$ & b)~$V/I_{\rm c}$ & c)~$Q/I_{\rm c}$ & d)~$U/I_{\rm c}$\\
\includegraphics[angle=-90,width=2.in]{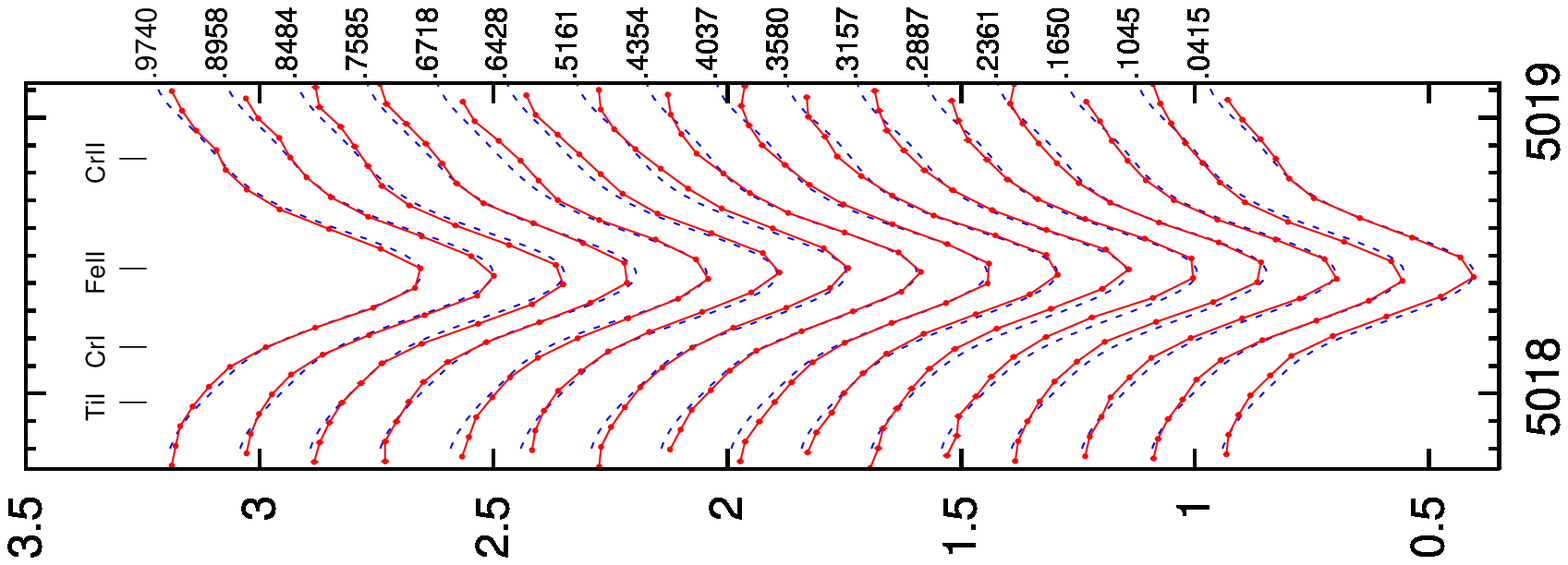} &
\includegraphics[angle=-90,width=2.in]{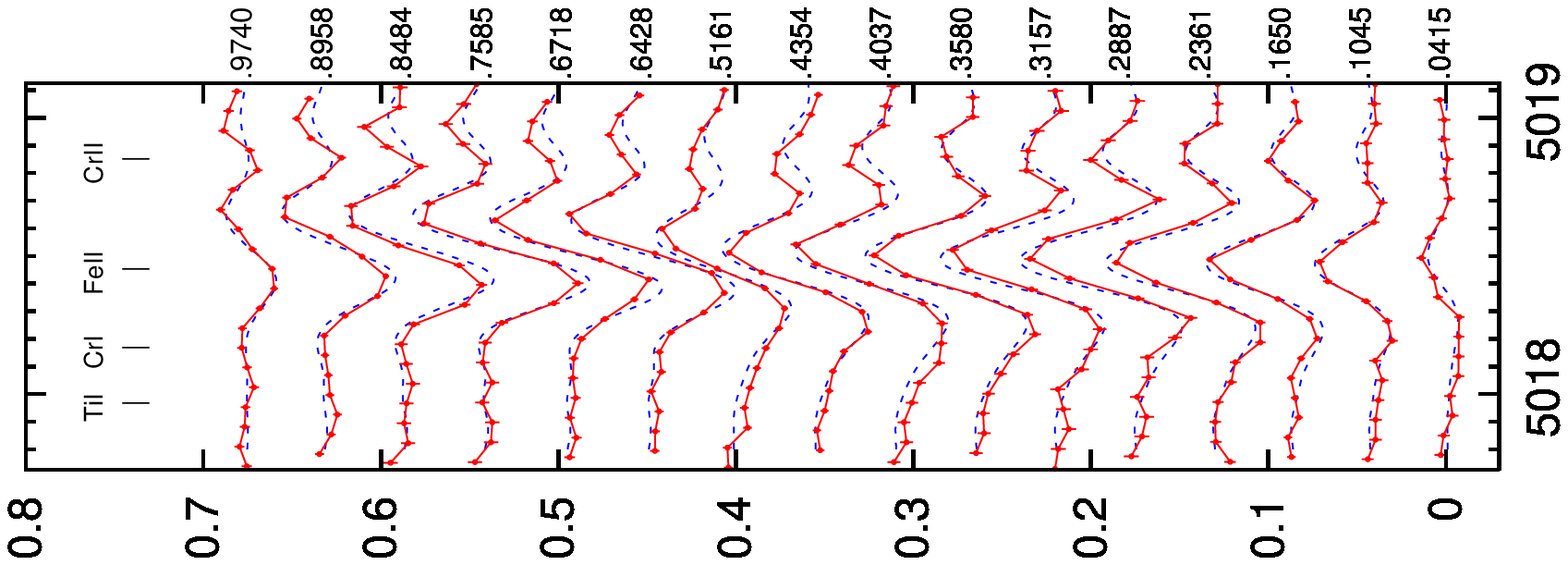} &
\includegraphics[angle=-90,width=2.in]{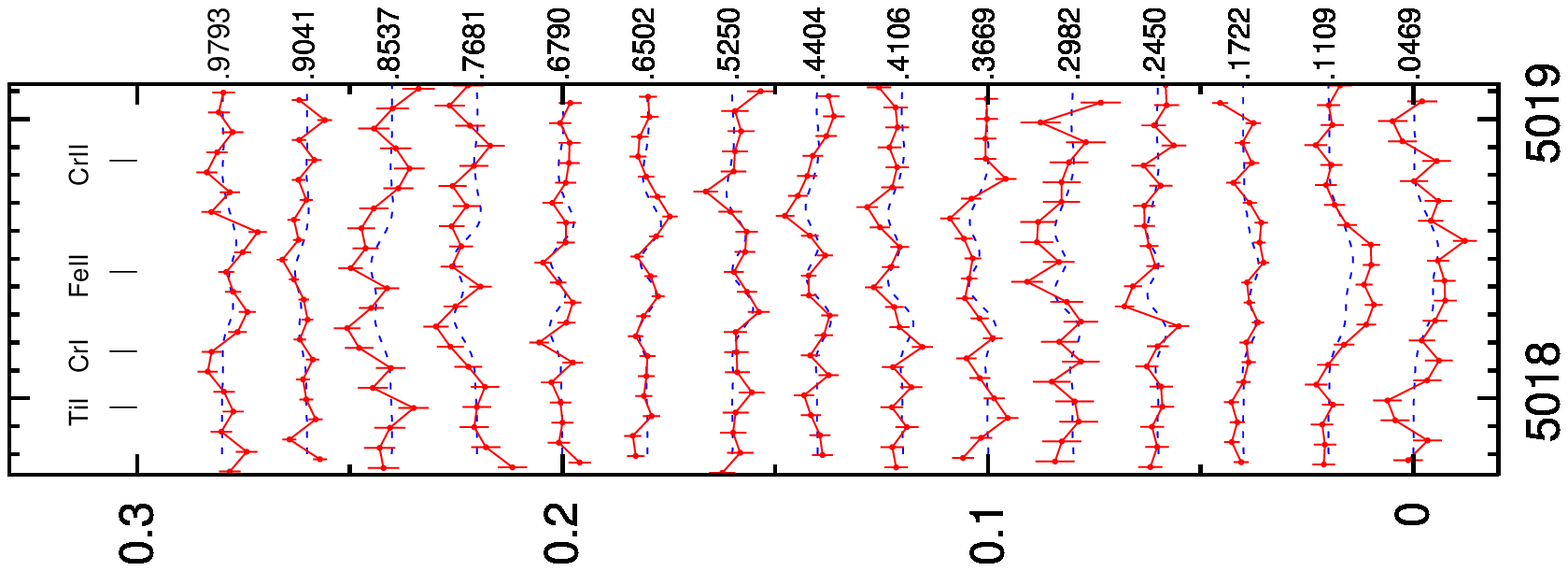} &
\includegraphics[angle=-90,width=2.in]{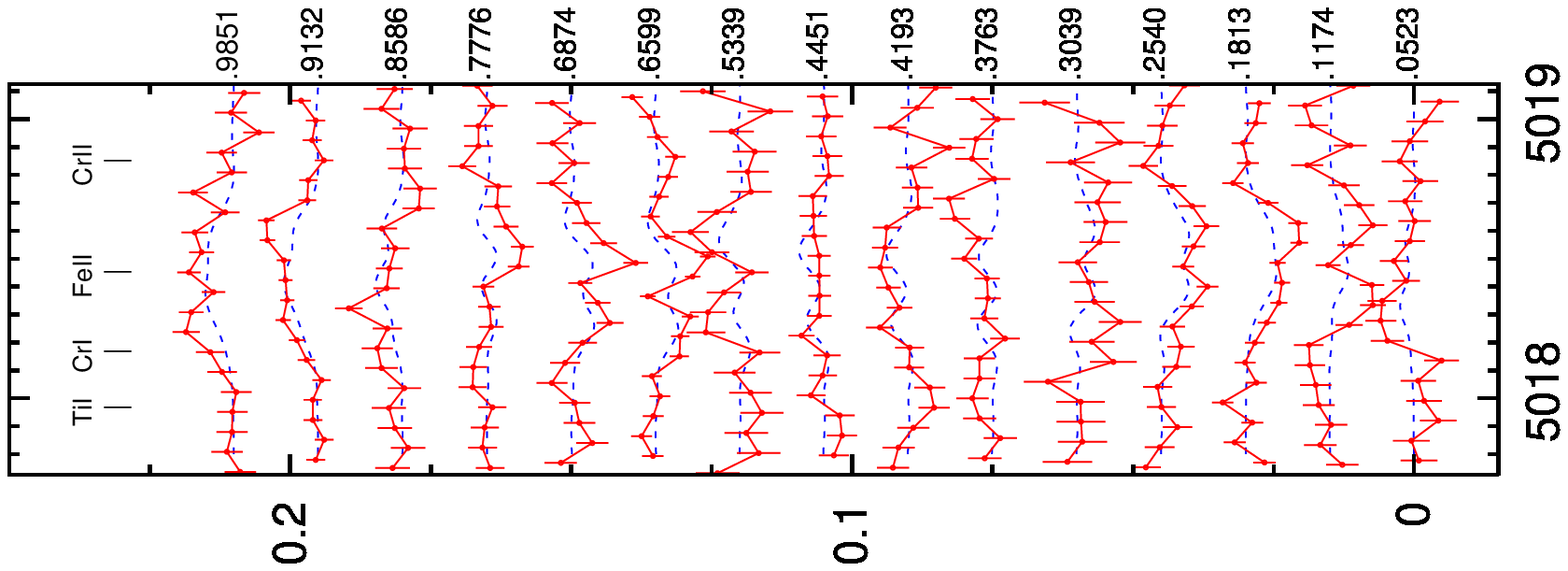}
\end{tabular}
}
\caption{Strong lines. Observed (solid lines) and simulated (dashed lines)
Stokes $I$ (a), $V$ (b), $Q$ (c) and $U$ (d) line profiles for Fe\,{\sc ii}
$\lambda$5018.44\AA\, line. 
The vertical bars show the size of the respective observational errors. The
observed and simulated profiles are shifted by 0.15, 0.045, 0.02 and on 0.015
along the vertical axes for Stokes $I$, $V$, $Q$ and $U$ profiles respectively.
}
\label{fe5018IVQU}
}
\end{figure*}

The other data in Table~\ref{fe2sm}
are derived from the free
model parameters (Khalack~et~al.~\cite{khalack+03}). Here $\beta$ defines the
angle between the magnetic dipole axis and the stellar rotation axis (the angle
exists if these two axes cross each other). The distance of the magnetic dipole
center from the center of the star and one-half of the magnetic dipole size are
represented by variables $a_{\rm 0}$ and $a$, respectively. The variables
($B_{\rm p}, \lambda_{\rm p}, \delta_{\rm p}$) and ($B_{\rm n}, \lambda_{\rm n},
\delta_{\rm n}$) specify the location of the positive and negative magnetic poles
at the stellar surface and the respective strength of the magnetic field.

\begin{figure}[th]
\includegraphics[angle=-90,width=3.5in]{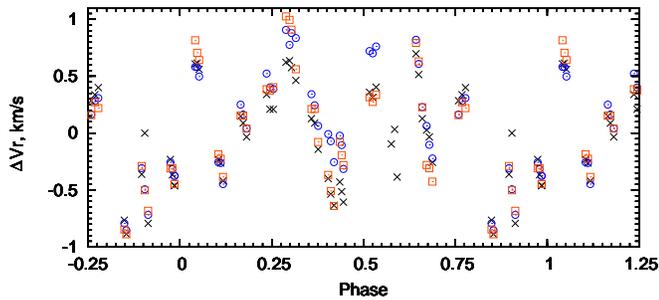}
\caption{The difference in radial velocity between the observed and
calculated Stokes $I$ profiles for the Fe\,{\sc ii} lines $\lambda$4923.927\AA\,
(crosses), $\lambda$5018.44\AA\, (open circles) and $\lambda$5169.033\AA\,
(open squares). The mean radial velocities $V_{\rm r}$=-7.82 km~$s^{-1}$,
-7.51 km~$s^{-1}$, -7.23 km~$s^{-1}$ obtained from the simulation of
Fe\,{\sc ii} $\lambda$4923.927\AA, $\lambda$5018.44\AA, $\lambda$5169.033\AA\,
lines respectively (see Table~\ref{fe2sm}), have been taken into account.
\label{VelocityDiff} }
\end{figure}

\subsubsection{The strong lines
\label{strong}}

The Fe\,{\sc ii} $\lambda$4923.927, $\lambda$5018.44
and $\lambda$5169.03 lines are the most
prominent Fe lines in the spectrum of 78 Vir, and show the clearest
Stokes $Q$ and $U$ signatures.
The first profile is formed essentially by the single line Fe\,{\sc ii}
$\lambda$4923.927 (see Table~\ref{tab3}) and appears to contain no
important blends. The agreement
between the observed and simulated data for Fe\,{\sc ii} $\lambda$4923.927
is similar to that shown in Fig.~\ref{fe5018IVQU} (for Fe~{\sc ii}~$\lambda 5018$).
Independent Stokes~$I$ spectra (each corresponding to a slightly different phase)
were obtained with each of the Stokes $V$, $Q$ and $U$ spectra, and
although only one set of Stokes $I$ profiles are presented in the figure, all
were taken into account during the calculation of the
$\chi_{\rm I}^2$-function (Eq.~\ref{chi2}). The best-fit values of
the Stokes $IVQU$ $\chi^2$-functions for the Fe\,{\sc ii} $\lambda$4923.927 line
are given in the third column of Table~\ref{fe2sm}.

\begin{figure*}[th]
\parbox[t]{\textwidth}{
\centerline{%
\begin{tabular}{@{\hspace{-0.22in}}c@{\hspace{-0.23in}}c@{\hspace{-0.23in}}c@{\hspace{-0.23in}}c}
a)~$I/I_{\rm c}$ & b)~$V/I_{\rm c}$ & c)~$Q/I_{\rm c}$ & d)~$U/I_{\rm c}$\\
\includegraphics[angle=-90,width=2.in]{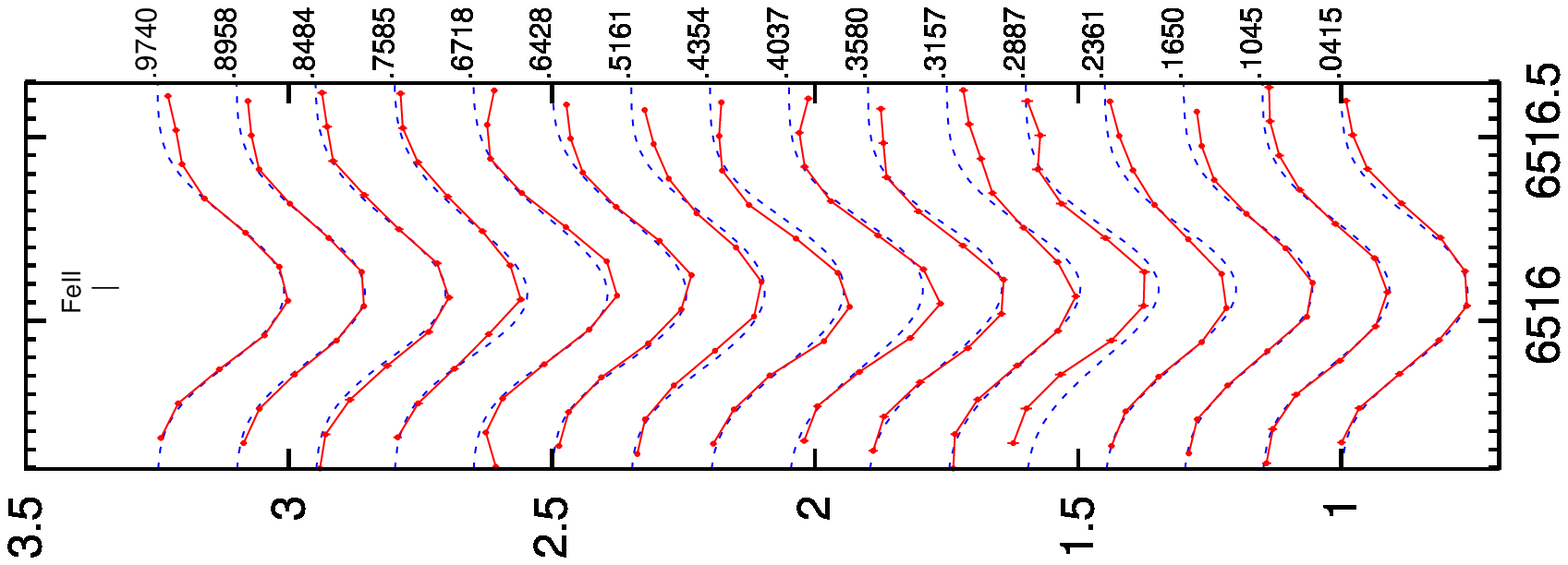} &
\includegraphics[angle=-90,width=2.in]{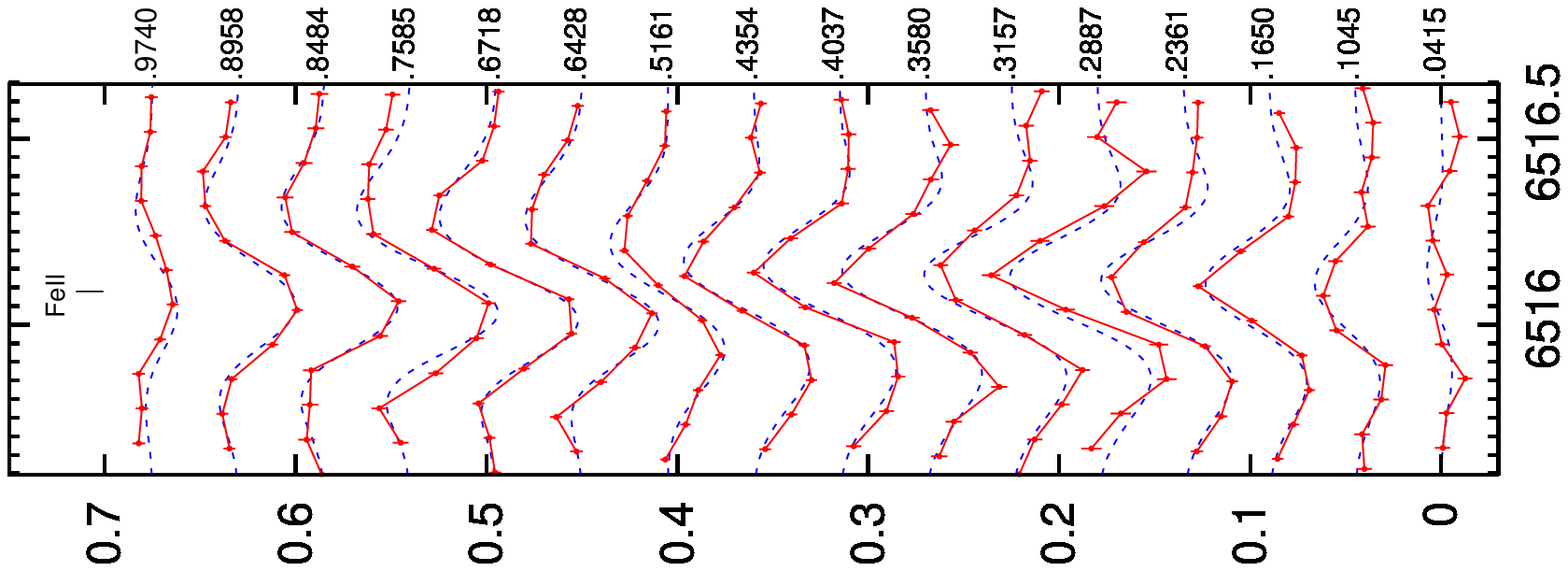} &
\includegraphics[angle=-90,width=2.in]{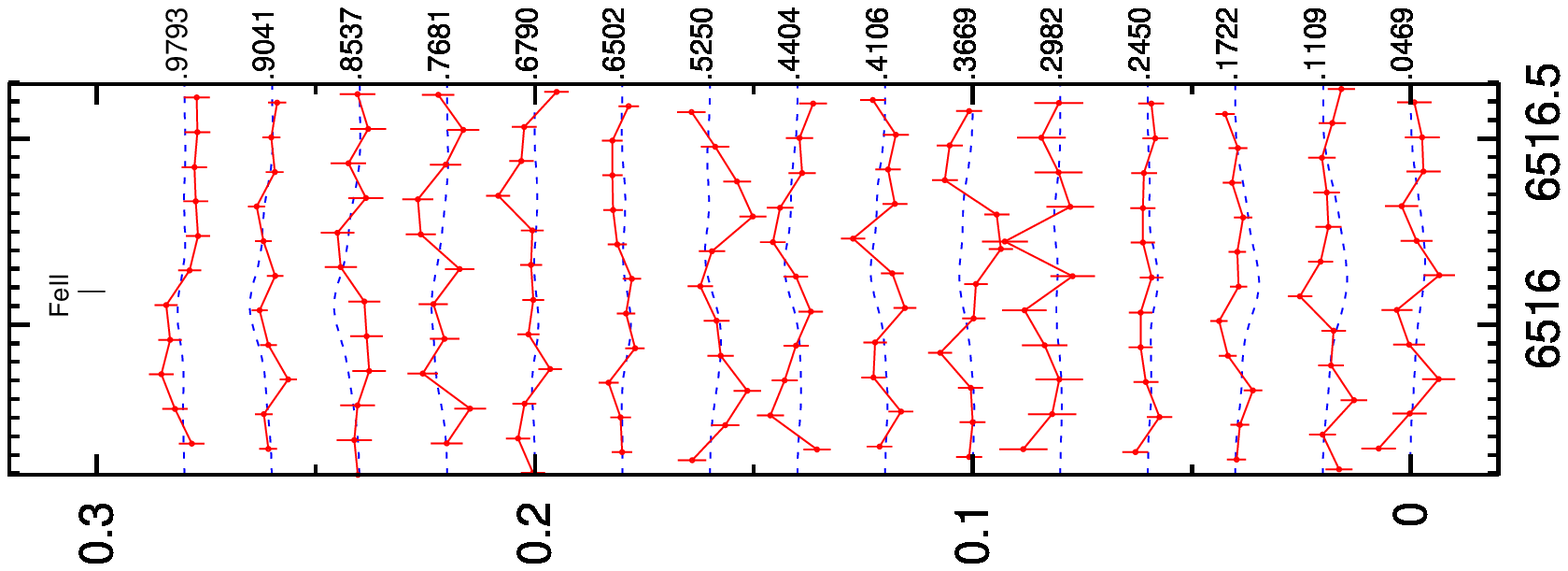} &
\includegraphics[angle=-90,width=2.in]{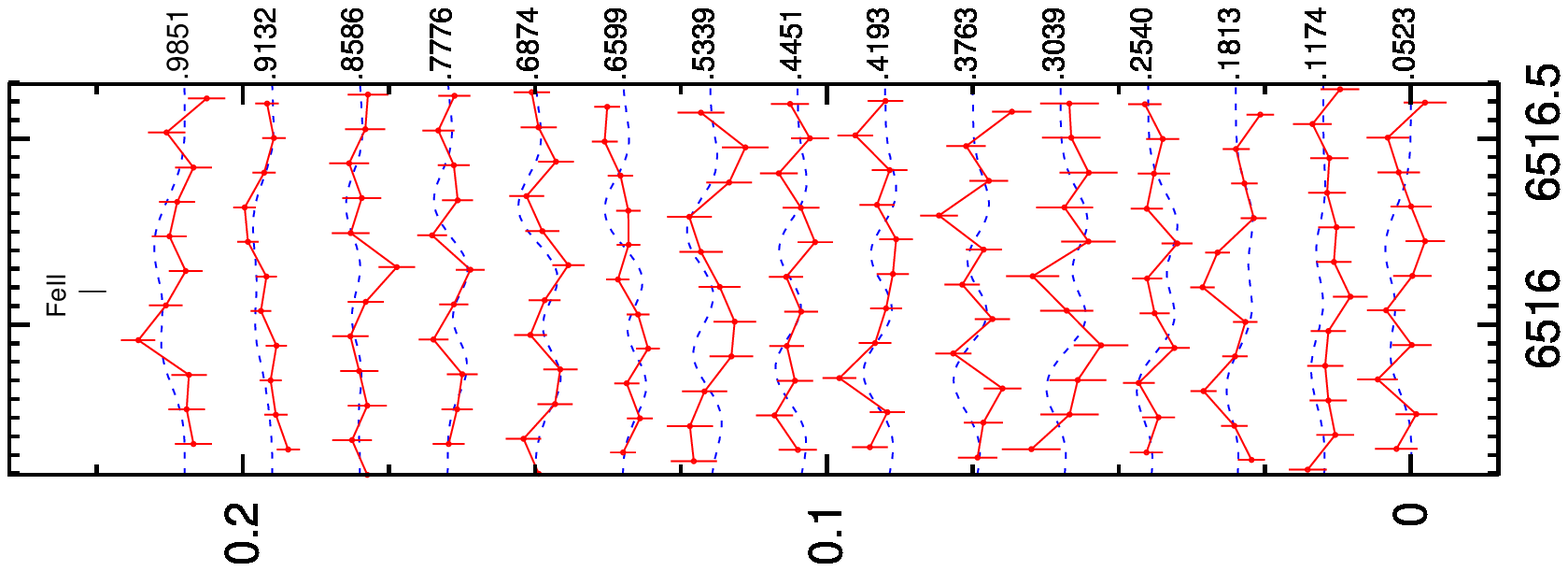}
\end{tabular}
}
\caption{The same as at the Fig.~\ref{fe5018IVQU}, but for Fe\,{\sc ii}
$\lambda$6516.08\AA.
\label{fe6516IVQU}}
}
\end{figure*}

The second line profile is composed mainly of the Fe\,{\sc ii} $\lambda$5018.44\AA\,
line, but is also contaminated by contributions from the Ti\,{\sc i}
$\lambda$5017.95, Cr\,{\sc i} $\lambda$5018.15 and
Cr\,{\sc ii} $\lambda$5018.84 lines. It seems that the
Cr\,{\sc ii} $\lambda$5018.84\AA\, is responsible for the
blend in the red wing of the Stokes~$I$ and $V$ profiles (Fig.~\ref{fe5018IVQU}).
Supposing a uniform Cr distribution,
this blend is
well fit for a Cr abundance of $\log Cr/N_{\rm tot}$=-3.38 dex.
The best fit $\chi^2$ values for the Stokes $IVQU$ profiles of Fe\,{\sc ii}
$\lambda$5018.44 are given in the fourth column of Table~\ref{fe2sm}.

In the case of Fe\,{\sc ii} $\lambda$5169.033, the profile is blended by
the weaker Fe\,{\sc i} $\lambda$5168.898 and $\lambda$5169.296 lines
(see Table~\ref{tab3}). According to Kochukhov et al.~(\cite{Kochukhov+04})
$\log gf$=-0.786 given in the GRIFON list for the Fe\,{\sc i} $\lambda$5169.296\AA\,
line is too high to match the solar spectrum, and they recommend to use a
decreased $\log gf$=-2.15. This value is employed here for the respective
profile simulation and provides much better agreement with the observed data.
The behaviour of the best fit Stokes~$IVQU$ profiles with stellar rotational
phase for this particular line is very similar to that shown in
Fig.~\ref{fe5018IVQU} for Fe\,{\sc ii} $\lambda$5018.44. The sixth column of
Table~\ref{fe2sm}
presents the best fit parameters and the
model characteristics for Fe\,{\sc ii} $\lambda$5169.033.


The concordance of the observed and simulated Stokes~$I$ and $V$ profile
variations as a function of stellar rotation is extremely good for these lines.
On the other hand, the fit to the
Stokes $Q$ and $U$ profiles is only approximate.
The simulated profiles show generally the same intensity,
qualitative structure and variability as the observations.
However, even with these relatively noisy data, it is clear
that the model does not reproduce the observations within the
errors at some phases (for example, see phases 0.1174, 0.5339 and 0.9851 in Fig. 2).



In order to verify the assumption of a uniform surface distribution of iron in
the atmosphere of 78 Vir, we have calculated the difference in radial velocity
between the observed and simulated Stokes $I$ profiles for the strong
Fe\,{\sc ii} lines $\lambda$4923.927, $\lambda$5018.44 and
$\lambda$5169.033\AA\, (see Fig.~\ref{VelocityDiff}).

The mean (averaged for all the available observational phases) radial
velocities obtained from the simulation are taken into account in the calculation
of the velocities of the respective simulated profiles.
Fig.~\ref{VelocityDiff} shows that the derived differences in $V_{\rm r}$
almost coincide for all three lines. A moderate disagreement may exist only
for the data obtained in the vicinity of rotational phase $\varphi$=0.5, when
the negative magnetic pole is most visible. The variation of radial
velocity, which shows a reasonably coherent variation with phase from
about -1 to +1~km/s, may reflect a mildly non-uniform distribution of Fe,
unmodelled structure in the magnetic field, or other unaccounted-for physical
processes in the stellar atmosphere.







\subsubsection{The moderate strength lines}

From the list of the Fe\,{\sc ii} lines selected for analysis (see Table~\ref{tab3})
the $\lambda$4620.52, $\lambda$5197.58,
$\lambda$6432.68 and $\lambda$6516.08 lines are weaker
than the $\lambda$4923.927, $\lambda$5018.44
and $\lambda$5169.03 lines.
Due to the weaker polarised signal, a few spectra
with comparatively large observational errors have been excluded from the simulation
(phases 0.5757, 0.5847 and 0.5914).

The line Fe\,{\sc ii} $\lambda$4620.521 is primarily responsible for the
formation of the observed profile. No blends were taken into account during the
simulation in this case. 
The Fe\,{\sc ii} $\lambda$5197.577 line is blended by the
weak Fe\,{\sc ii} $\lambda$5197.48 line, but provides the main contribution
to the observed profile. The behaviour of the best fit Stokes~$IVQU$ profiles with
stellar rotational phase for these lines is very similar to that
shown in Fig.~\ref{fe6516IVQU} for Fe\,{\sc ii} $\lambda$4620.521\AA. The first
and seventh columns of Table~\ref{fe2sm} 
present the best fit parameters and the model characteristics for Fe\,{\sc ii}
lines $\lambda$4620.521\AA\, and $\lambda$5197.577\AA, respectively.

The Fe\,{\sc ii} $\lambda$6432.68\AA\, line is also essentially unblended. The
agreement between the observed and simulated Stokes~$IVQU$ data for this
particular line is very similar to that presented by
Fig.~\ref{fe6516IVQU}. The Fe\,{\sc ii} $\lambda$6516.08\AA\, line is the
primary contributor to the formation of the corresponding observed
Stokes~$IVQU$ profiles. The agreement of the best-fit simulation with the
observed data for Fe\,{\sc ii} $\lambda$6516.08\AA\, line is shown at
Fig.~\ref{fe6516IVQU}. The tenth and eleventh columns at the Tables~\ref{fe2sm}
contain the best fit parameters and model characteristics for the
Fe\,{\sc ii} $\lambda$6432.68\AA\, and $\lambda$6516.08\AA\, lines,
respectively. The best fit of the simulated Stokes~$IVQU$ profiles for the
Fe\,{\sc ii} $\lambda$6516.08\AA\, line is statistically better than that
obtained for Fe\,{\sc ii} $\lambda$6432.68\AA.

\subsubsection{Separate analysis of 
the Stokes $I$ and $V$ spectra}

The other Fe\,{\sc ii} lines $\lambda$4635.316\AA, the $\lambda$5100\AA-group,
$\lambda$5362.87\AA\, and $\lambda$6247.56\AA\, show no variability
in the Stokes~$Q$ and $U$ spectra and hence only the Stokes~$I$ and $V$ spectra
are taken into account during the simulation. These lines are selected for the
analysis in order to check the resulting magnetic field structure of 78 Vir on
the basis of a more thorough line list.

\begin{figure}[th]
\centerline{%
\begin{tabular}{@{\hspace{-0.22in}}c@{\hspace{-0.25in}}c}
a)~$I/I_{\rm c}$ & b)~$V/I_{\rm c}$ \\
\includegraphics[angle=-90,width=2.in]{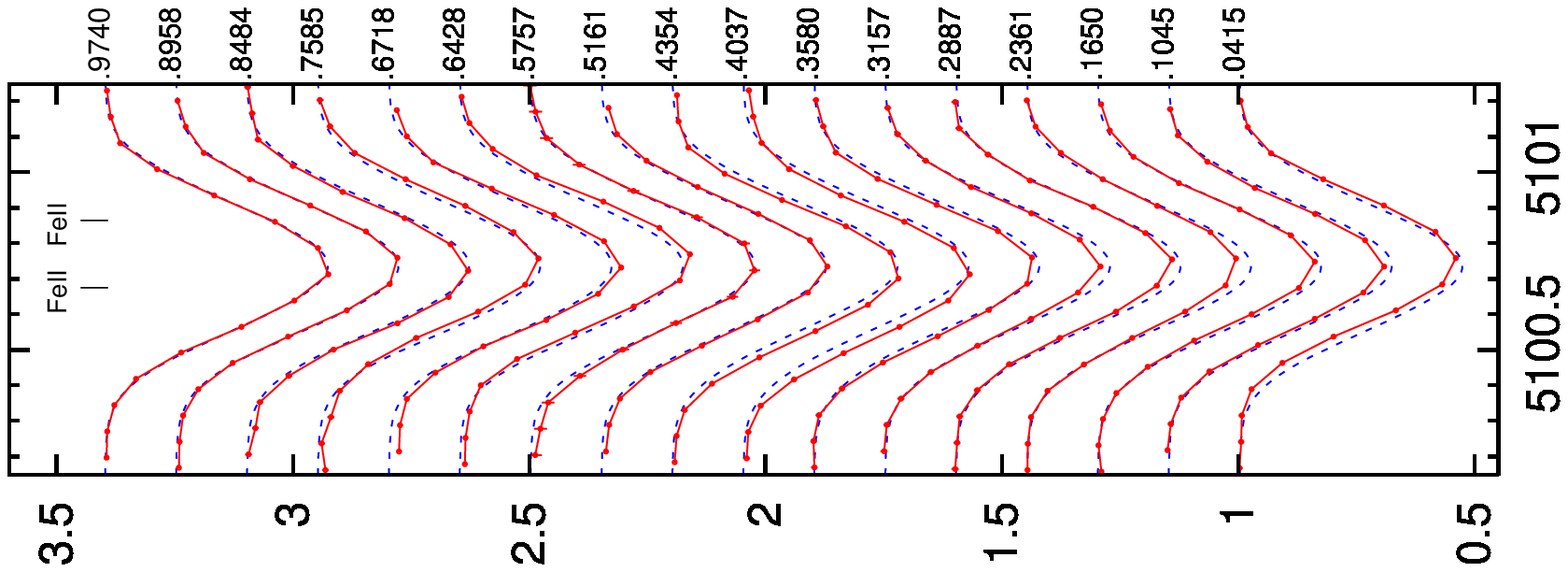} &
\includegraphics[angle=-90,width=2.in]{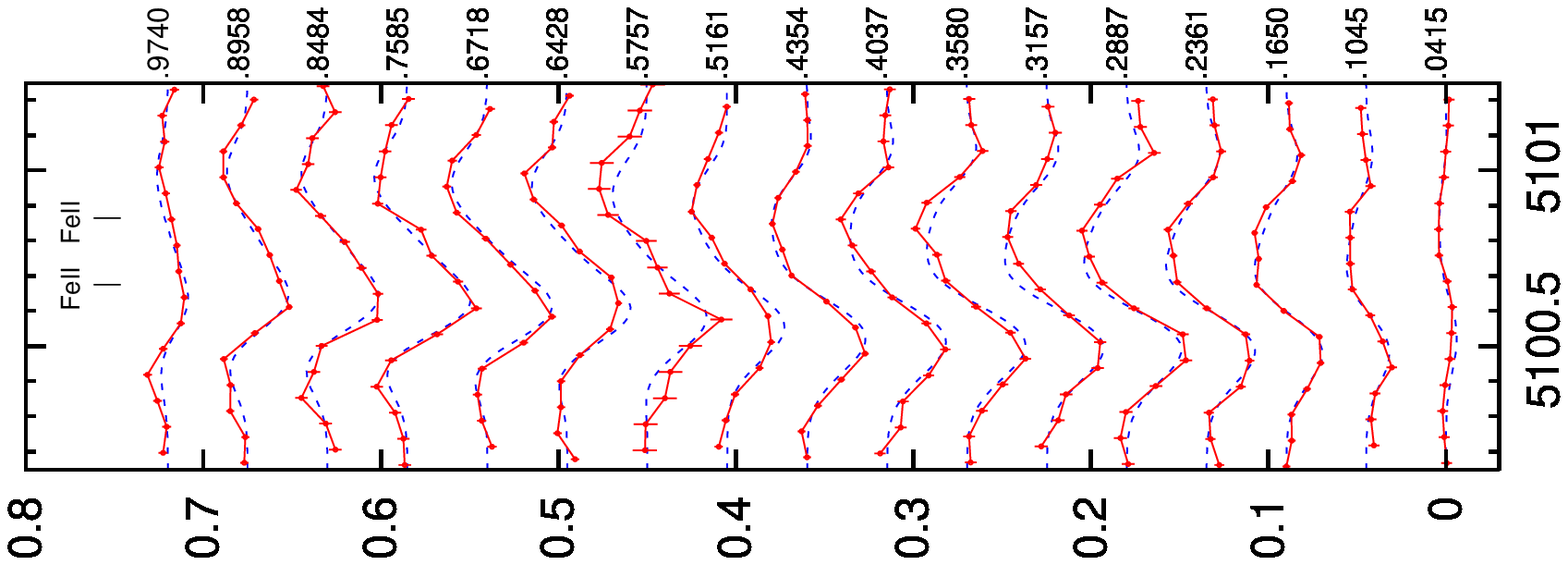}
\end{tabular}}
\caption{Observed (solid lines) and simulated (dashed lines)
Stokes I (a) and V (b) profiles for Fe\,{\sc ii} $\lambda$5100.607\AA,
$\lambda$5100.664\AA, $\lambda$5100.727\AA\, and $\lambda$5100.852\AA\,
lines. 
The vertical bars show the size of respective observational errors. The
observed and simulated profiles are shifted on 0.15, and on 0.045 along the
vertical axes for Stokes $I$ and $V$ profiles correspondingly.
\label{fe5100IV} }
\end{figure}


The Fe\,{\sc ii} $\lambda$4635.316 line is the main contributor to the
formation of the corresponding observed Stokes~$I$ and $V$ profiles, although
it is blended by the comparatively weak Fe\,{\sc i} $\lambda$4635.846\AA\,
line. The agreement between the observed and simulated data for this particular
line is similar to that shown in Fig.~\ref{fe5100IV}. The superposition of the
Fe\,{\sc ii} $\lambda$5100.607, $\lambda$5100.664, $\lambda$5100.727
and $\lambda$5100.852 lines is responsible for the formation of Stokes~$I$
and $V$ profiles at $\sim 5100.7$~\AA\ (see Table~\ref{tab3}).
Fig.~\ref{fe5100IV} shows the simulated Stokes~$I$ and $V$ profiles, which fit
well the observed profiles (see also the respective $\chi^2$ value in the fifth
column of Table~\ref{fe2sm}).

In the region of the Fe\,{\sc ii} $\lambda$5362.87 line the observed
Stokes~$I$ and $V$ profiles are formed mainly by this line and by a weak
contribution from the Fe\,{\sc ii} $\lambda$5362.74, $\lambda$5362.98,
Cr\,{\sc i} $\lambda$5362.87 and Cr\,{\sc ii} $\lambda$5363.88 lines.
The Fe\,{\sc ii} $\lambda$6247.557 line also
provides the main contribution to the observed Stokes~$I$ and $V$ profiles and is
blended by the Fe\,{\sc ii} $\lambda$6247.35 line.
The best fit simulated data show almost the same fit quality as is
shown in Fig.~\ref{fe5100IV}.
Simulation of the Stokes~$I$
and $V$ profiles in the regions of these lines results in almost the same
configuration of the magnetic field structure as obtained from the analysis of the
stronger Fe\,{\sc ii} lines, for which all four Stokes parameters
were taken into account during the simulation.

\begin{figure}[th]
\includegraphics[angle=-90,width=3.5in]{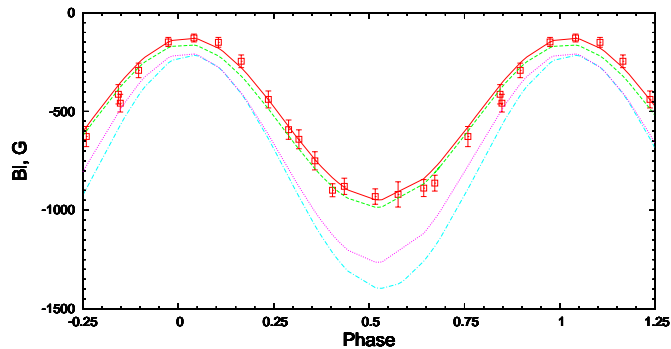}
\caption{Longitudinal magnetic field data with phase
obtained from analysis of Fe\,{\sc ii} $\lambda$6432.68 (solid line),
Fe\,{\sc ii} $\lambda$5018.44 (dashed line), Fe\,{\sc ii} $\lambda$6516.08
(dotted line) and Fe\,{\sc ii} $\lambda$5197.58 (dash-dotted line).
The open squares with $1\sigma$ error bars correspond to the
LSD data published by Wade et al. (\cite{Wade+00b}).
\label{Blon1} }
\end{figure}


\subsection{Integral magnetic field characteristics}
\label{integral}

In order to check the agreement of the derived magnetic field model with other
available magnetic field data for 78 Vir, we calculate the intensity-weighted,
averaged (over the visible stellar disk) longitudinal magnetic field $B_{\rm l}$
and the normalized equivalent widths of the Stokes~$Q$ and $U$ profiles for all
the analysed phases. The normalization procedure is performed in accordance
with the method described by Wade~et~al. (\cite{Wade+00b}) over the passband of
the analysed line profiles.

All the available longitudinal magnetic field measurements for 78 Vir are
plotted in Fig.~\ref{Blon1} of Leone~\&~Catanzaro~(\cite{L+C01}). This figure shows
the good agreement of the $B_{\rm l}$ measurements obtained by
Wade~et~al.~(\cite{Wade+00b}) using the Least-Squares Deconvolution (LSD)
technique (Donati~et~al.~\cite{Donati+97}) with the majority of other
observational data. Therefore, in this paper we just compare our results with
the LSD longitudinal field data.

As demonstrated by Wade~et~al. (\cite{Wade+00b}), the 78 Vir Stokes~$Q$ and $U$
LSD equivalent widths are proportional to the BBLP measured at that phase
(Leroy~\cite{Leroy95}) with a line scaling factor. This factor might be
different for the analysed lines (see the last column in Table~~\ref{tab3}),
but is the same for the both Stokes~$Q$ and $U$ equivalent widths for a
particular spectral line.

\begin{figure}[th]
\includegraphics[angle=-90,width=3.5in]{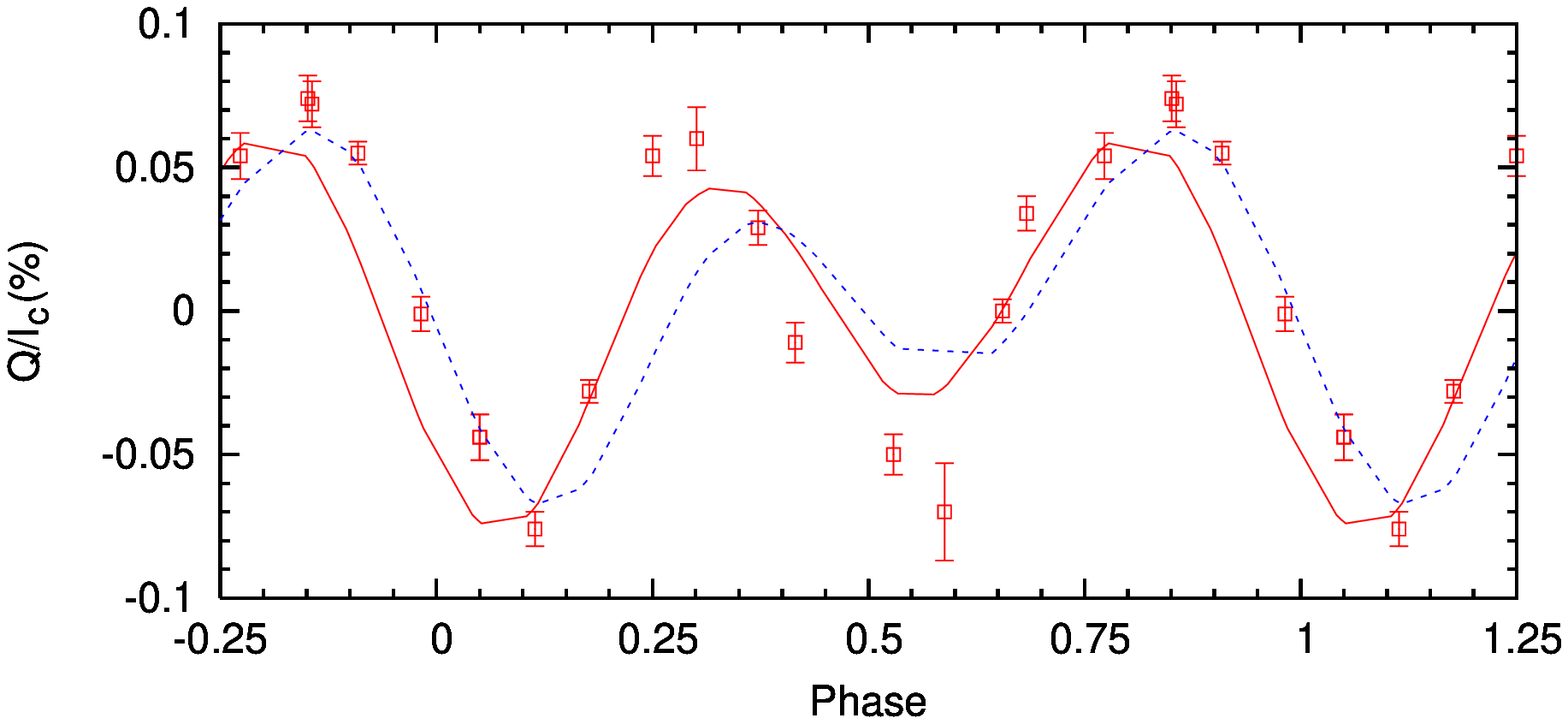}
\includegraphics[angle=-90,width=3.5in]{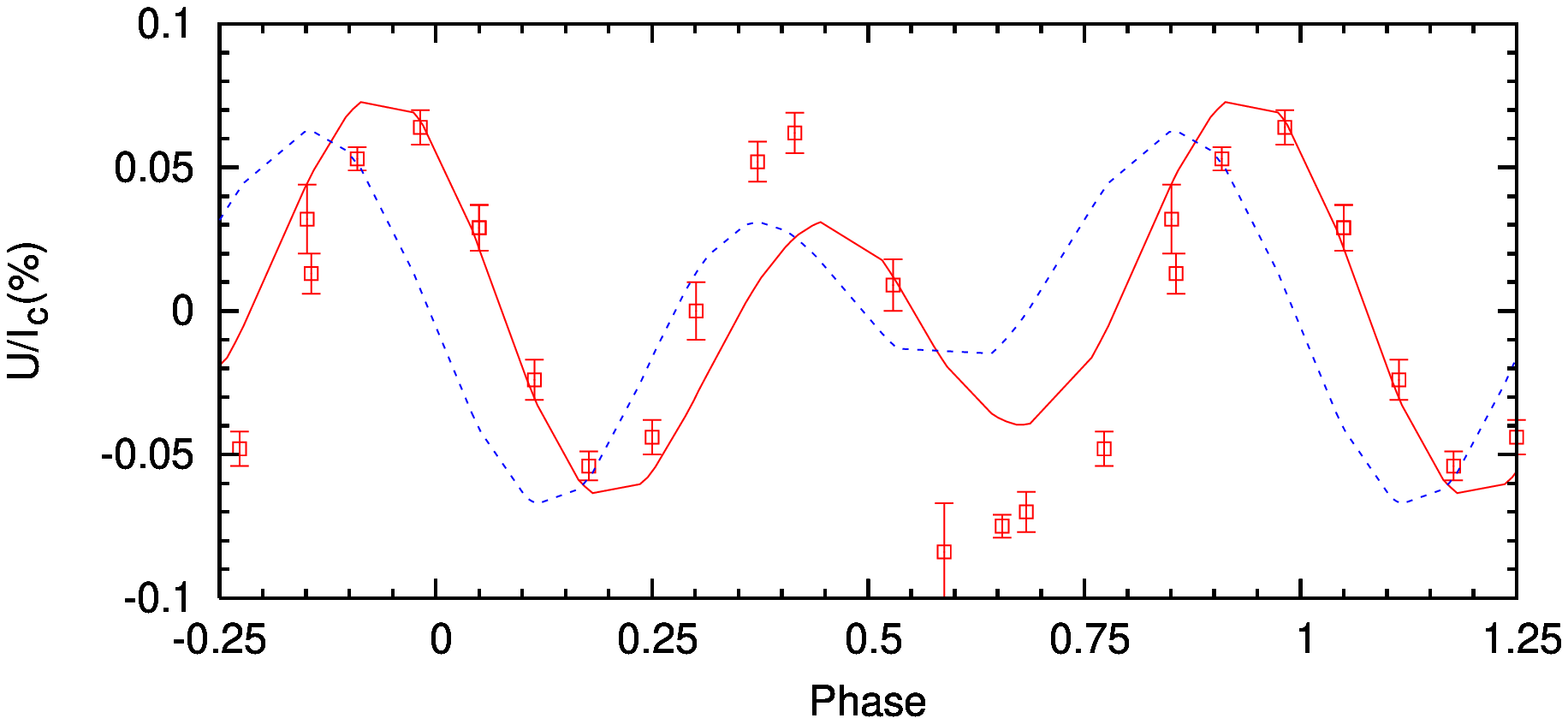}
\caption{Comparison of scaled net linear polarization data obtained
from the best fit simulations of Fe\,{\sc ii} $\lambda$4923.927 (solid line)
and Fe~{\sc ii} $\lambda$6516 (dashed line), and from LSD profiles (Wade~et~al.~\cite{Wade+00b})
(open squares) for 78 Vir with respective $1\sigma$ error bars.
\label{Wade-4923} }
\end{figure}

Neither the $B_{\rm l}$ measurements nor the BBLP data are included directly
into the model minimization procedure. However, given that the model
successfully reproduces the Stokes profiles with which these quantities are
fundamentally related, we should expect that the model magnetic field
configuration is capable of reproducing them. Using the same stellar atmosphere
model, the longitudinal field calculated from the model fits to
the strongest Fe\,{\sc ii} line $\lambda$5018.44\AA\, and to
the weak Fe\,{\sc ii} line $\lambda$6432.68\AA\, show a
good agreement with the LSD $B_{\rm l}$ data (see Fig.~\ref{Blon1}). The
longitudinal fields calculated from the simulation results for the other
Fe\,{\sc ii} lines appear to be shifted downward in longitudinal
field intensity relative to the LSD $B_l$ variation (see Fig.~\ref{Blon1}).

\begin{table*}[th]
\parbox[t]{3.5in}{
\center{\caption[]{The same as at Table~\ref{fe2sm},
but for Cr\,{\sc ii} and Ti\,{\sc ii} lines}
\label{crti}}
\vspace{0.in}
\begin{tabular}{l|ccccccc|ccc}
\hline\hline
Line         & Cr\,{\sc ii}& Cr\,{\sc ii}& Cr\,{\sc ii}& Cr\,{\sc ii}& Cr\,{\sc ii}&Cr\,{\sc ii}&
 & Ti\,{\sc ii}& Ti\,{\sc ii}& \\
 \AA             & 4592 & 4634 & 5237 & 5310 & 5407 & 5421 & $\sigma_{\rm er}$ & 5188 & 5336 &
 $\sigma_{\rm er}$ \\
\hline
$\chi^2_{\rm I}$ & 14.82& 13.12& 10.51&  6.47&  6.20&  9.48&     & 21.53& 12.45& \\
$4\chi^2_{\rm V}$&  8.83&  6.39& 12.70&  8.64& 11.11& 11.50&     & 24.28&  9.30& \\
$6\chi^2_{\rm Q}$&  5.62&  6.18&   -  &  8.18&  8.55& 11.76&     &  8.77&  5.79& \\
$6\chi^2_{\rm U}$&  6.48&  5.88&   -  &  6.63&  9.17& 12.33&     &  7.45&  6.45& \\
$\chi^2_{\rm w}$ &  8.95&  7.89& 11.60&  7.49&  8.76& 11.27&     & 15.51&  8.50& \\
$\chi^2$         &  4.76&  4.18&  6.84&  2.77&  2.98&  4.09&     &  7.58&  4.20& \\
\hline
Qr, kG           &  175 &  186 &  185 &  182 &  161 &  171 &  30 &  287 &  271 &  40 \\
$a_{\rm 1}, 10^{-3}$&10.2& 9.5 &  9.7 &  8.9 &  9.6 &  8.6 & 1.5 &  3.7 &  3.8 & 1.4 \\
$\lambda_{\rm 1}$& 19\dr& 15\dr& 16\dr& 25\dr& 17\dr& 21\dr& 5\dr& 28\dr& 31\dr&17\dr\\
$\delta_{\rm 1}$ &-41\dr&-42\dr&-45\dr&-49\dr&-40\dr&-46\dr& 5\dr&-39\dr&-41\dr&19\dr\\
$a_{\rm 2}, 10^{-3}$&4.4&  4.1 &  3.8 &  3.9 &  3.6 &  3.7 & 0.3 &  2.8 &  3.5 & 1.0 \\
$\lambda_{\rm 2}$&103\dr&120\dr&111\dr&115\dr&109\dr&120\dr& 8\dr&113\dr&113\dr&19\dr\\
$\delta_{\rm 2}$ &-17\dr&-18\dr&-17\dr&-12\dr& -2\dr&-19\dr&10\dr& -8\dr&-14\dr&12\dr\\
$\Omega$         &109\dr&117\dr&   -  &123\dr&115\dr&109\dr&14\dr& 97\dr& 90\dr&19\dr\\
$i$              & 26\dr& 26\dr& 26\dr& 26\dr& 22\dr& 27\dr& 5\dr& 25\dr& 18\dr& 5\dr\\
$\log(El/N_{\rm tot})$
                 &-3.90 & -4.02& -4.04& -3.91& -4.18& -4.17& 0.19& -5.48& -5.68& 0.22\\
$V_{\rm r}$      & -8.25& -8.02& -8.80& -8.41& -8.05& -8.25& 1.23& -5.61& -7.91& 1.26\\
$V_{\rm e}\sin{i}$&11.7 & 10.8 & 11.8 & 11.5 & 10.9 & 12.2 & 1.1 & 13.4 & 13.3 & 1.5 \\
\hline
$\beta$          &123\dr&119\dr&126\dr&131\dr&127\dr&124\dr& 5\dr&117\dr&112\dr& 5\dr\\
$a_{\rm 0}, 10^{-3}$&6.1&  5.2 &  5.5 &  5.2 &  5.1 &  4.9 & 1.4 &  2.5 &  2.9 & 1.7 \\
$a, 10^{-3}$     &  5.0 &  5.1 &  5.0 &  4.6 &  5.1 &  4.4 & 1.4 &  2.1 &  2.2 & 1.7 \\
$B_{\rm p}$, kG  & 3.52 & 3.86 & 3.71 & 3.36 & 3.32 & 3.05 & 0.4 & 2.46 & 2.41 & 0.5 \\
$\lambda_{\rm p}$&-11\dr&-10\dr&-11\dr& -8\dr& -8\dr& -7\dr& 8\dr&-18\dr&-23\dr& 8\dr\\
$\delta_{\rm p}$ &-33\dr&-29\dr&-36\dr&-41\dr&-37\dr&-34\dr&10\dr&-27\dr&-22\dr&11\dr\\
$B_{\rm n}$, kG  & -3.44& -3.78& -3.62& -3.29& -3.24& -2.99& 0.4 & -2.45& -2.41& 0.5 \\
$\lambda_{\rm n}$&169\dr&170\dr&169\dr&172\dr&171\dr&173\dr& 8\dr&161\dr&156\dr& 8\dr\\
$\delta_{\rm n}$ & 33\dr& 29\dr& 35\dr& 40\dr& 37\dr& 34\dr&10\dr& 27\dr& 22\dr&11\dr\\
\hline\hline
\end{tabular}
}
\end{table*}

\begin{figure*}[th]
\parbox[t]{\textwidth}{
\centerline{%
\begin{tabular}{@{\hspace{-0.22in}}c@{\hspace{-0.23in}}c@{\hspace{-0.23in}}c@{\hspace{-0.23in}}c}
a)~$I/I_{\rm c}$ & b)~$V/I_{\rm c}$ & c)~$Q/I_{\rm c}$ & d)~$U/I_{\rm c}$\\
\includegraphics[angle=-90,width=2.in]{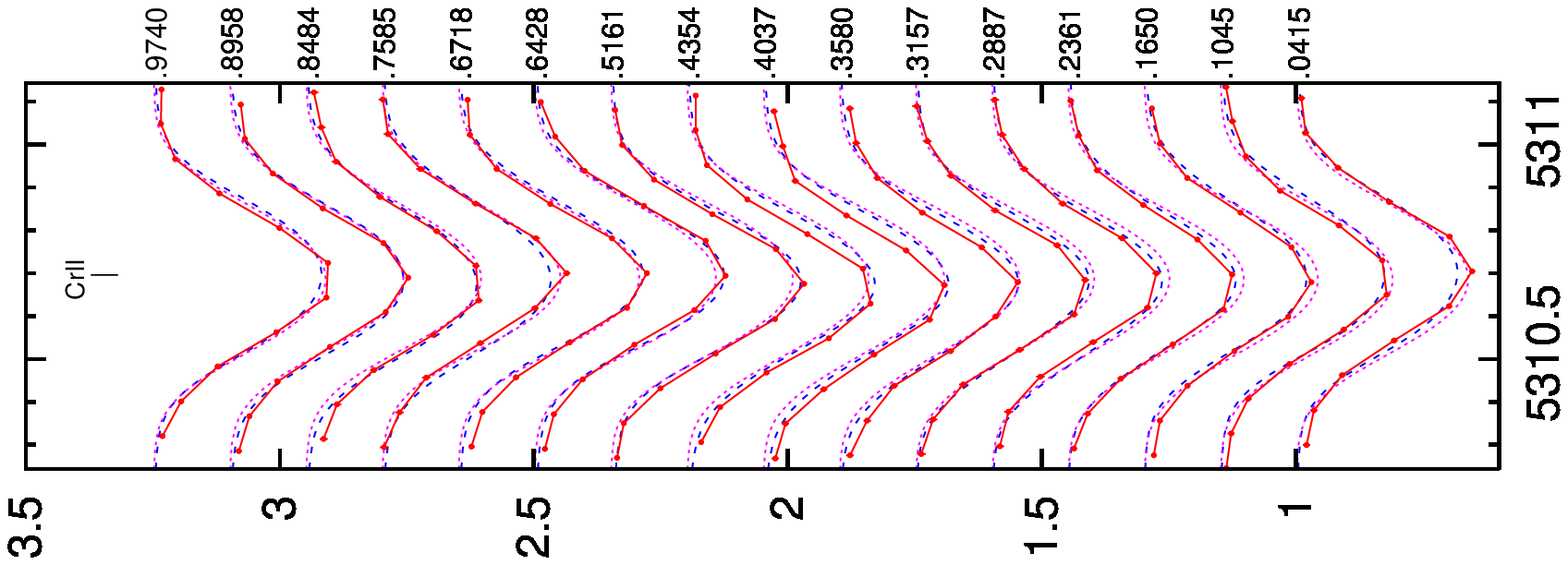} &
\includegraphics[angle=-90,width=2.in]{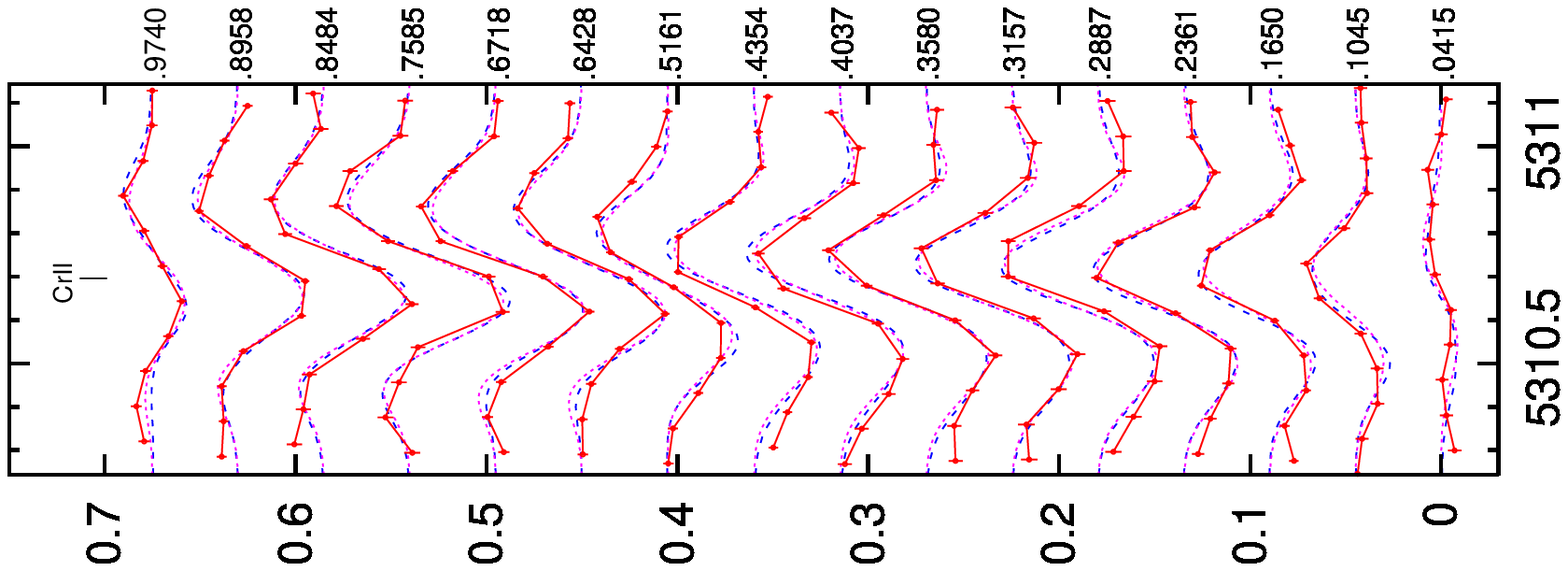} &
\includegraphics[angle=-90,width=2.in]{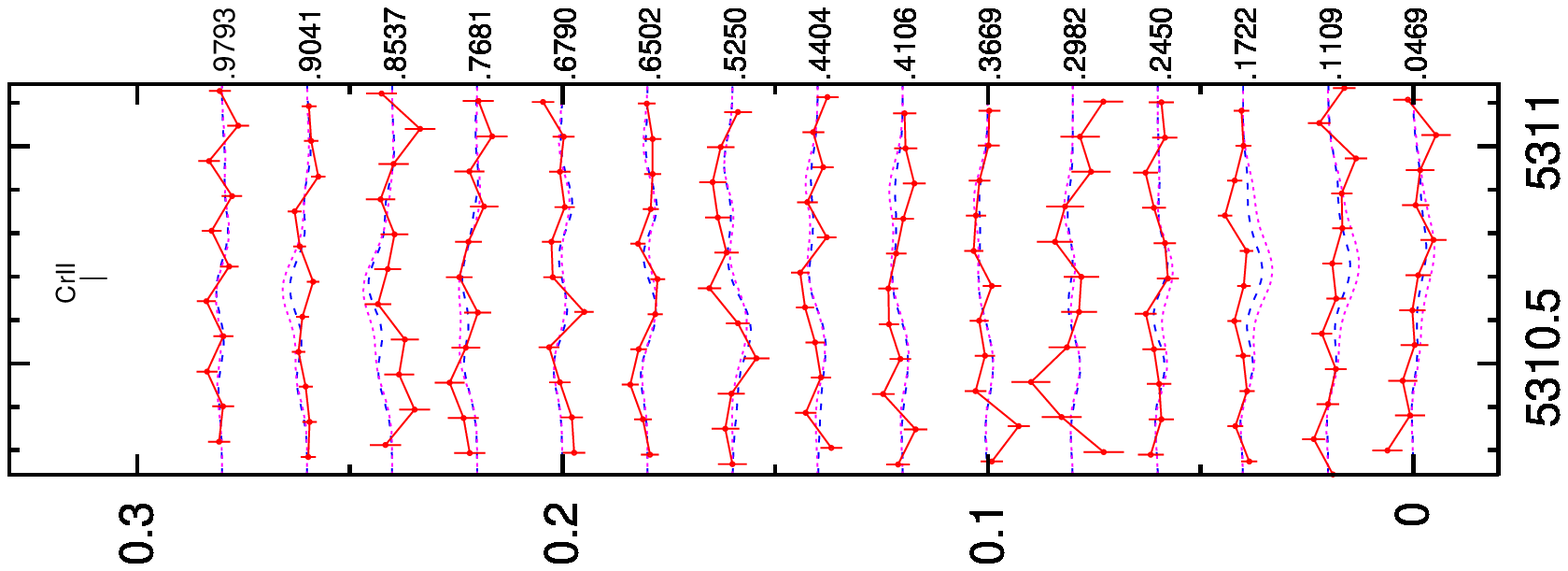} &
\includegraphics[angle=-90,width=2.in]{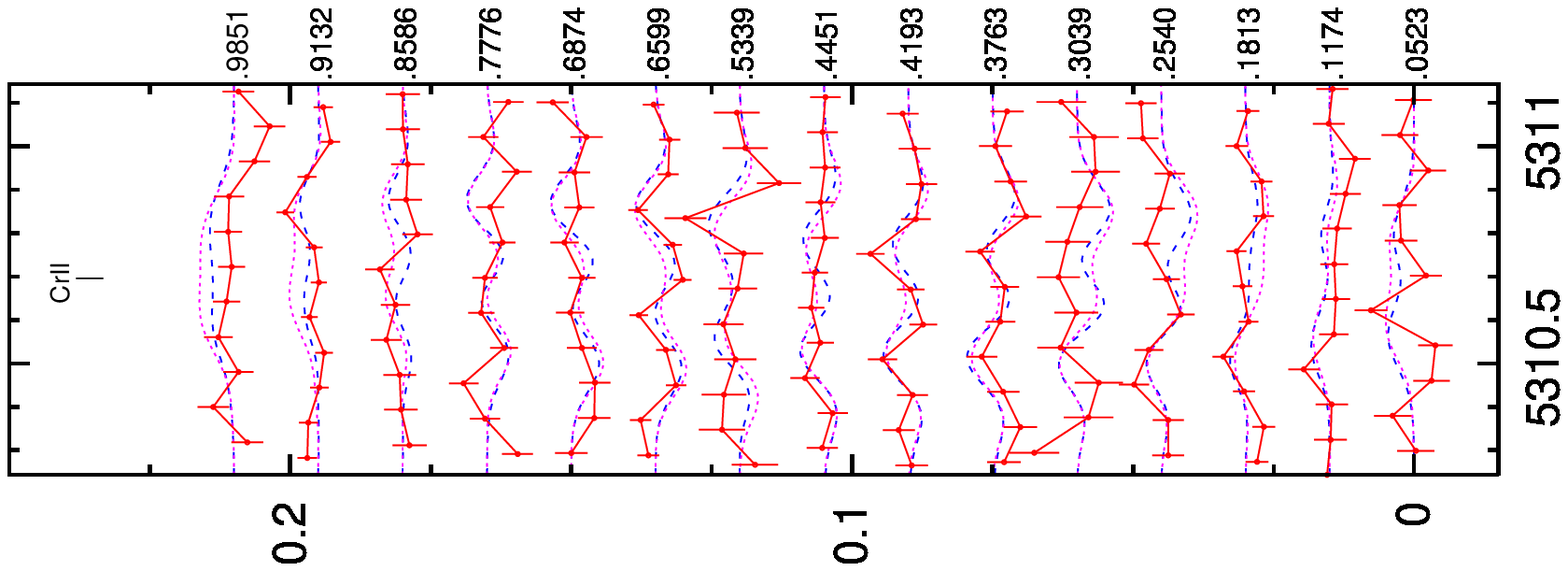}
\end{tabular}
}
\caption{The same as at the Fig.~\ref{fe5018IVQU}, but for Cr\,{\sc ii}
$\lambda$5310.687\AA. Non-uniform chromium distribution in the stellar
atmosphere is assumed during this simulation (dashed line). For comparison
we show also the best fit simulation result ($\chi^2=4.24$) for uniform chromium
distribution (dotted line).
\label{Cr5310IVQU}}
}
\end{figure*}

The calculated Stokes~$Q$ and $U$ equivalent widths for all the analysed line
profiles are multiplied by the respective line scaling factor in order to fit
them to the LSD Stokes~$Q$ and $U$ data. For each analysed line a scaling
factor is determined with the help of the Least Squares method. The respective
results are given in Table~\ref{tab3}, taking into account that the LSD
equivalent widths of Stokes~$Q$ and $U$ profiles have a line scaling factor of 0.1
(Wade~et~al.~\cite{Wade+00b}) from comparison with the BBLP data
(Leroy~\cite{Leroy95}). Fig.~\ref{Wade-4923} presents the Stokes $Q$ and $U$
equivalent widths variability  with phase, derived
from the best fit simulations of Fe\,{\sc ii} $\lambda$4923.927
(with $\Omega=109\degr$) and Fe\,{\sc ii} $\lambda$6516.08 (with $\Omega=128\degr$).
They are scaled by a factor of 0.08 (for $\lambda 4923$) and 0.03 (for $\lambda 6516$)
and plotted over the
Stokes $Q$ and $U$ equivalent widths derived from LSD profiles. It is remarkable
that the simulated polarimetric data for Fe\,{\sc ii} $\lambda$4923.927\AA\,
and $\lambda$5018.44\AA\, lines have almost
the same scaling factor as the LSD Stokes $Q$ and $U$ equivalent widths.
Theoretically, the line scaling factor depends on the Land\'e factor and on the
degree of line saturation (Wade~et~al.~\cite{Wade+00a}). Table~\ref{tab3}
shows that in general the lines with a high mean Land\'e factor have also a
comparatively high line scaling factor. Nevertheless, this dependence is not
clearly pronounced by the sample of derived data. This is because, for some
lines, this effect is reduced by blend contamination of the analysed
Stokes $Q$ and $U$ profiles
(for example the Fe\,{\sc ii} $\lambda$5100 line group,
$\lambda$5197.03 and $\lambda$5362.87 lines) or by weak Stokes $Q$
and $U$ profile variability,
which results in a comparatively low precision of the calculated equivalent widths.

The LSD net linear polarisation (in the same way as the BBLP)
reflect the
contributions of many spectral lines with significant polarization. These lines
belong to a number of chemical elements, which apparently have non-uniform and
usually quite different abundance distributions. Such a composition of polarised
features results in LSD (or BBLP) data that may differ substantially from the
Stokes $Q$ and $U$ equivalent widths derived for a
particular Fe\,{\sc ii} line (see Fig.~\ref{Wade-4923}).
This is illustrated by the rather large differences between
the two simulated curves in Fig.~\ref{Wade-4923}. Moreover, lines of elements
with non-uniform abundance distributions can provide substantially different values
of the sky-projected position angle of the stellar rotation axis (see
Sect.~\ref{other}), and can exhibit totally different net linear polarisation
variations (e.g. 180$\degr$ out of phase with those shown here). The spectral
resolution and the observational errors of
the available spectra do not allow us to map confidently the detailed
distribution of chemical abundances nor the local patterns of the surface
magnetic field. The uncertainties in the details of the abundance distributions
and the magnetic field structure limit the precision of $\Omega$ as well as the
quality of simulated Stokes $Q$ and $U$ equivalent widths. Given these limitations,
we would characterise the agreement of the observed and simulated variations shown
in Fig.~\ref{Wade-4923} as very acceptable.

\begin{figure*}[th]
\parbox[t]{\textwidth}{
\centerline{%
\begin{tabular}{@{\hspace{-0.22in}}c@{\hspace{-0.23in}}c@{\hspace{-0.23in}}c@{\hspace{-0.23in}}c}
a)~$I/I_{\rm c}$ & b)~$V/I_{\rm c}$ & c)~$Q/I_{\rm c}$ & d)~$U/I_{\rm c}$\\
\includegraphics[angle=-90,width=2.in]{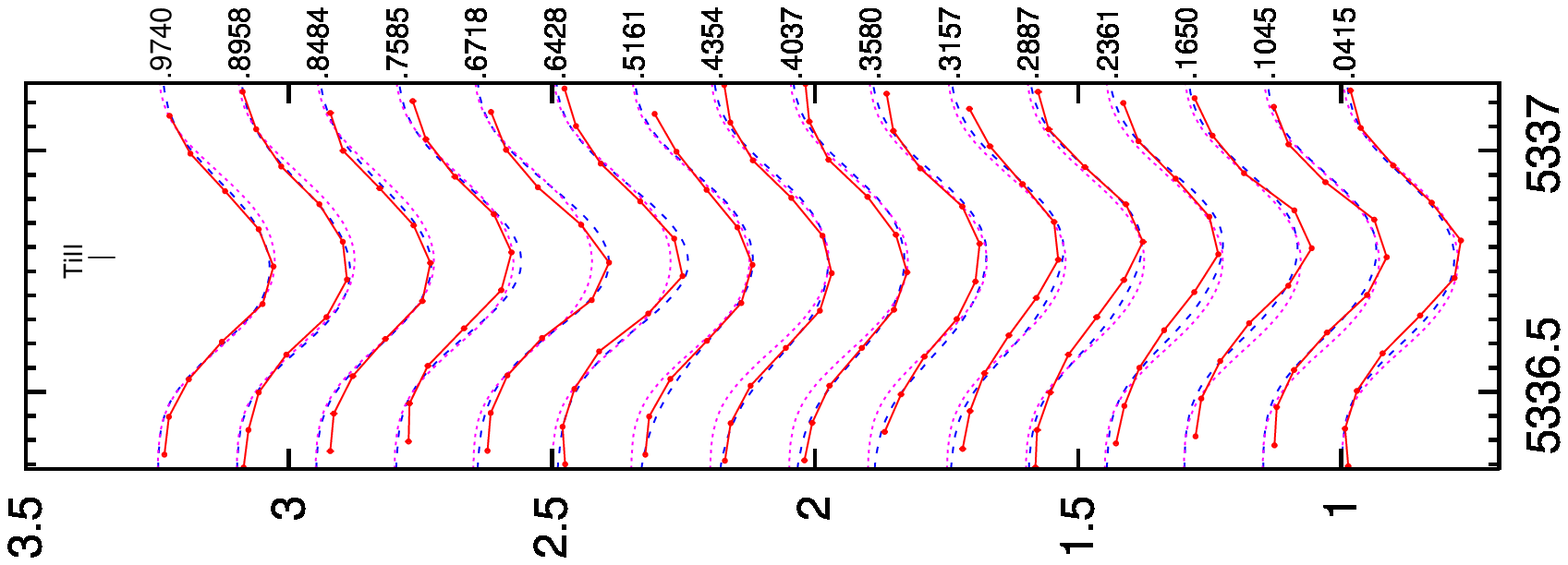} &
\includegraphics[angle=-90,width=2.in]{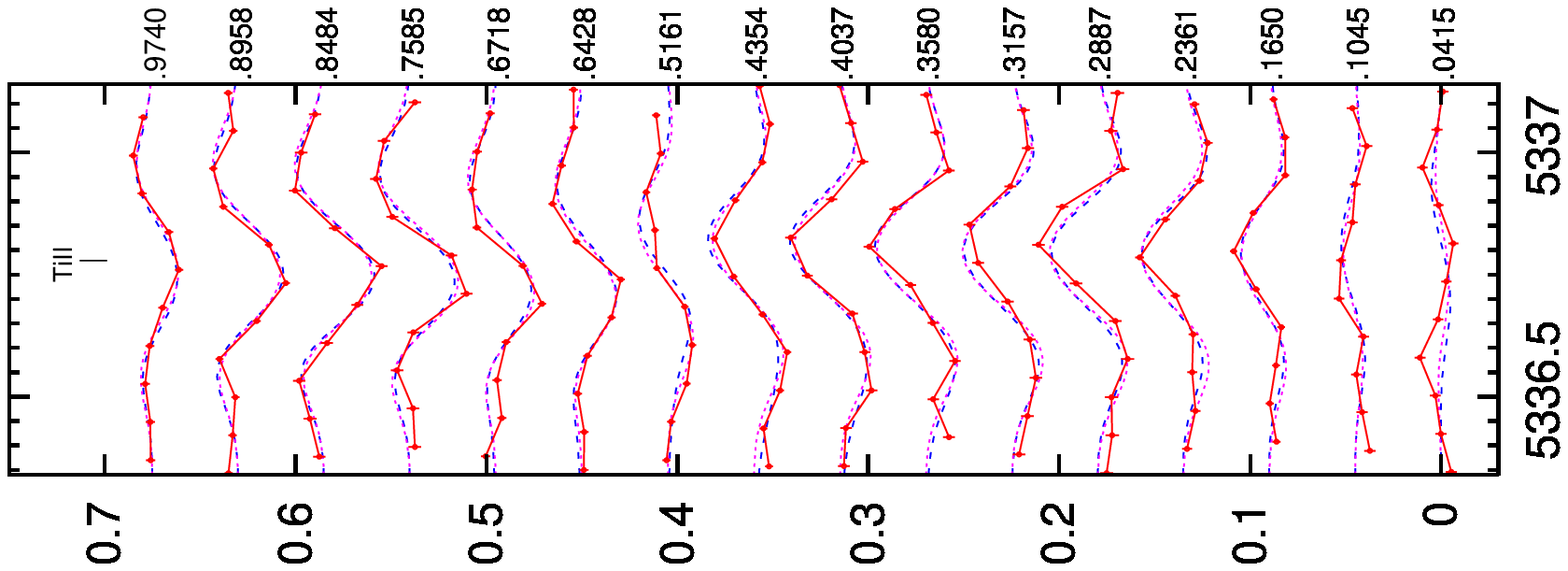} &
\includegraphics[angle=-90,width=2.in]{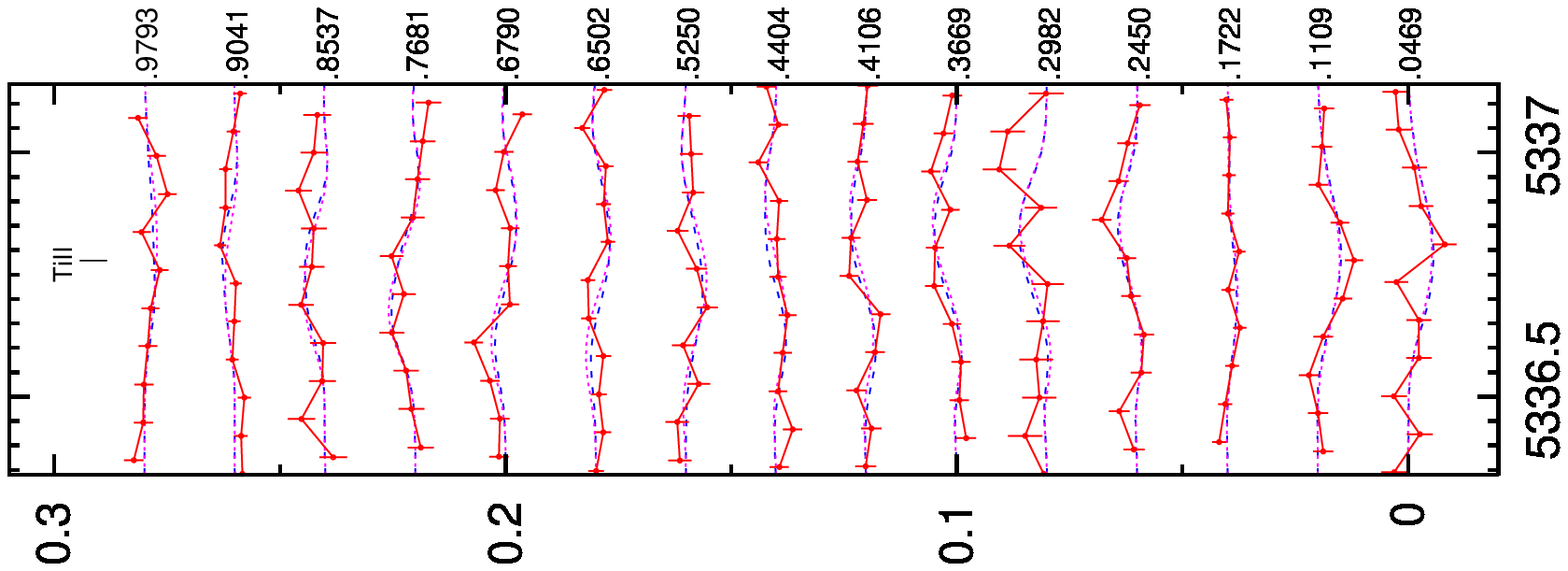} &
\includegraphics[angle=-90,width=2.in]{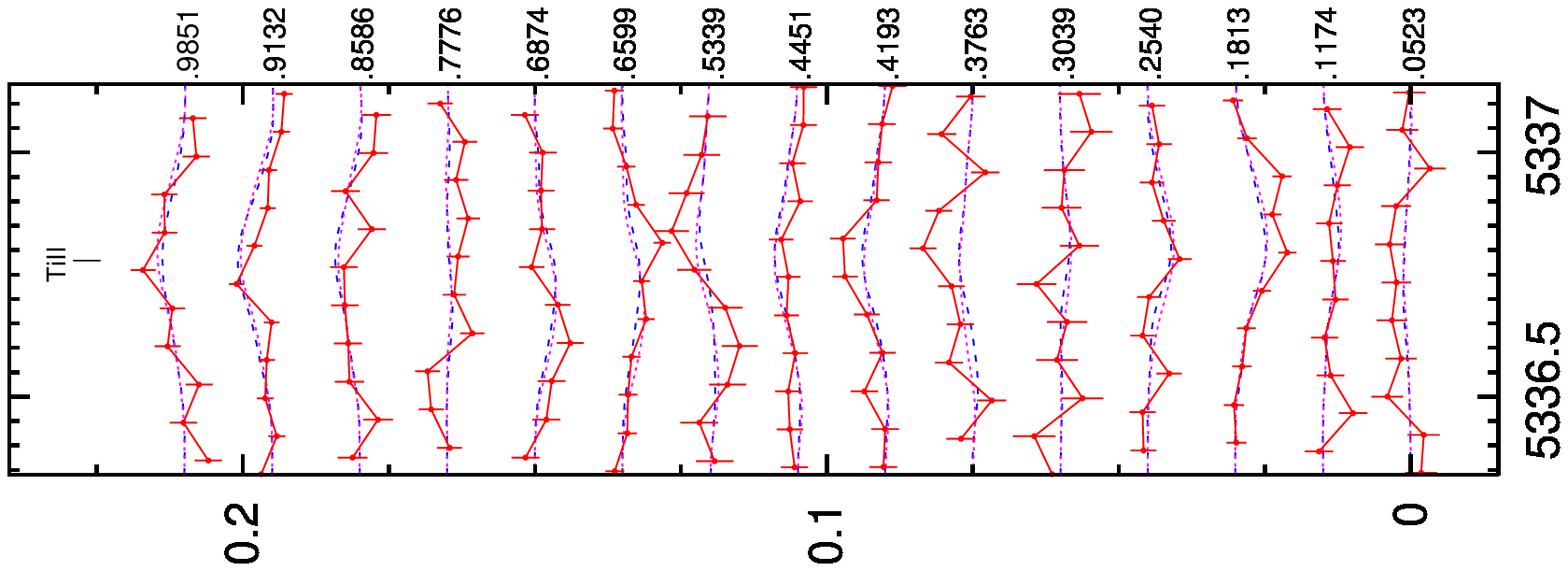}
\end{tabular}
}
\caption{The same as at the Fig.~\ref{fe5018IVQU}, but for Ti\,{\sc ii}
$\lambda$5336.771\AA. Non-uniform titanium distribution in the stellar
atmosphere is assumed during this simulation (dashed line).  For comparison
we show also the best fit simulation result ($\chi^2=4.58$) for a uniform titanium
distribution (dotted line).
\label{Ti5336IVQU}}
}
\end{figure*}

\subsection{Lines of other elements
\label{other}}




78 Vir is a chromium-rich star (Cowley~et~al.~\cite{C2J2}) and has some strong
Cr\,{\sc ii} lines, which are located in the observed spectral range. The most
prominent of them, Cr\,{\sc ii} $\lambda$4592.049, $\lambda$4634.07,
$\lambda$5237.329, $\lambda$5310.687,
$\lambda$5407.604 and $\lambda$5420.922\AA\,, have been selected for simulation.
Atomic parameters of spectral lines were extracted from the VALD database
(Kupka~et~al.~\cite{Kupka+99}) and from Raassen~\&~Uylings~(\cite{R+U98})
({\it ftp://ftp.wins.uva.nl/pub/orth}) in the case of Cr\,{\sc ii} lines.
These lines are not significantly blended by lines of other elements and
show variability of the Stokes $IV$ parameters with
phase. Some show marginally-detected Stokes $QU$ signatures as well.

We analyse also the two strong Ti\,{\sc ii} $\lambda$5188.68
and $\lambda$5336.771 lines.
These Ti lines are not significantly blended (although we take into account
a small contribution of the V\,{\sc i} $\lambda$5188.885 blend to the
profile formed by Ti\,{\sc ii} $\lambda$5188.68).

Table~\ref{crti} presents the results of the best fit simulations for the
Cr and Ti lines. 
The best fit quality is on average similar to that
obtained for the Fe\,{\sc ii} lines. The Stokes $I$ profiles for these particular
lines vary weakly with rotational phase, suggesting the presence of non-uniform
distributions of Cr\,{\sc ii} and Ti\,{\sc ii} in the atmosphere of 78 Vir.
To take into account the non-uniform abundance distributions the stellar
surface is divided into 30 areas, each characterised by an independent local
abundance of the analysed chemical element. The local abundances
are included as free parameters in the simulation procedure and the model
operates in this case with 41 parameters.
The best fit simulations of the aforementioned lines tentatively suggest that
chromium is enhanced in the vicinity of the negative magnetic pole,
while titanium is underabundant in this region (see Fig.~\ref{Abun}). The improvement
in the fit to Cr~{\sc ii} $\lambda 5310$ resulting from a non-uniform abundance distribution
is highly significant (reduction of $\chi^2_I$ by over 40\%), although the improvement for
Ti~{\sc ii} $\lambda 5336$ is much less so (just 10\%).
The averaged chromium and titanium abundances are given in Table~\ref{crti}.








\begin{figure*}[th]
\begin{tabular}{@{\hspace{+0.04in}}c@{\hspace{-0.05in}}c}
a) & b) \\
\includegraphics[scale=1.0]{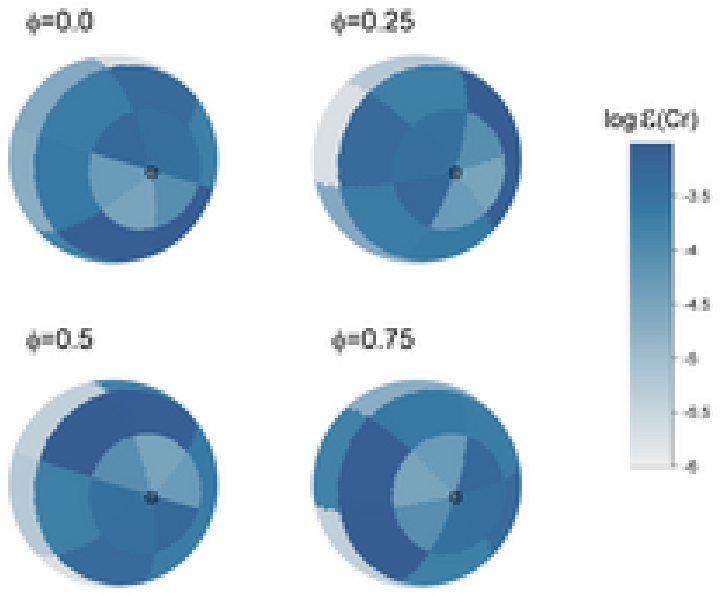} &
\includegraphics[scale=1.0]{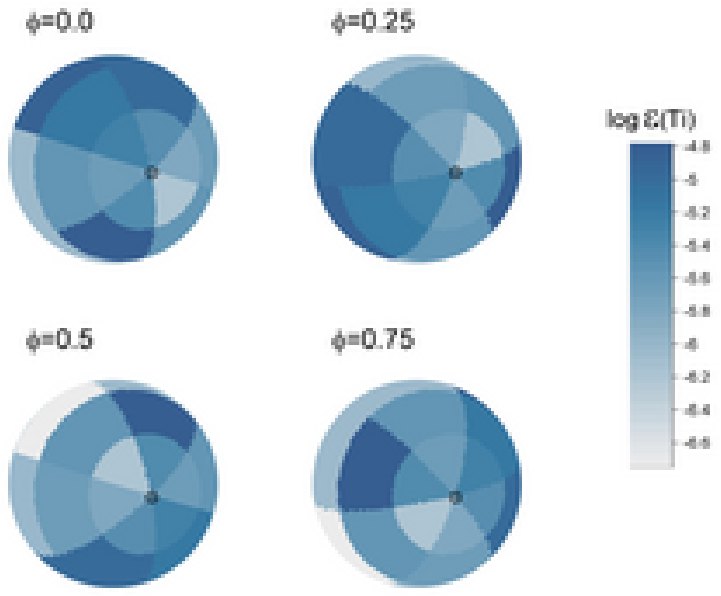}
\end{tabular}
\caption{Illustration of the a) chromium and b) titanium abundance
distribution in the stellar atmosphere of 78 Vir. Location of the
positive rotation pole is specified by open circle.
The stellar image has been rotated in the plane of the
page by the positional angle $\Omega$=110\dr.
\label{Abun} }
\end{figure*}

It is remarkable that the best fit simulations of Cr\,{\sc ii} lines
provide almost the same global magnetic field structure and the
sky-projected position angle of the stellar rotation axis as the Fe lines
(see Tables~\ref{fe2sm},~\ref{crti}). Meanwhile, the Ti\,{\sc ii} simulation
results in somewhat lower values of $\Omega$ and $\beta$.
These differences may results from the
non-uniform Ti\,{\sc ii} distribution in the stellar atmosphere.

\section{Discussion}

\label{discuss}

The abundance of Fe\,{\sc ii} derived in this study is $\log{Fe/N_{\rm tot}}=-3.16\pm 0.20$
(see Table~\ref{fe2sm}).
%
%
From analysis of spectra of 78 Vir in the regions
3850-3870~\AA\ and 3868-4650~\AA\
Adelman~(\cite{Adelman73b}) obtained an Fe\,{\sc ii} abundance of
$\log{Fe/N_{\rm H}}=-2.89$dex.
The differences in the iron abundances
{may be due to the inclusion/exclusion of polarized transfer,
microturbulence, effects of Balmer line wings, etc.}, as well as from
the different effective temperatures that have been employed for the
abundance analysis (for 78 Vir Adelman~(\cite{Adelman73a}) applied
$T_{\rm eff}=9950$K, significantly higher than the temperatures employed
in our model. 

For the 11 different Fe features studied here, we observe no systematic difference
in strength between weak and strong lines. Therefore, the results of this study are
consistent with the absence of important Fe stratification in the atmosphere of 78~Vir.

According to our results the global magnetic field structure of 78
Vir is well-described by a slightly decentered magnetic dipole.
Usually the starting point of a simulation (in the free parameter
hyperspace) is chosen to correspond to a central magnetic dipole,
but for all the analysed lines the final model results in a
slightly decentered magnetic dipole. This fact is in good
qualitative agreement with the results previously obtained by
Borra (\cite{Borra80}). In that paper Borra considered the
classical decentered dipole model, where the magnetic dipole size
is insignificant in comparison with the stellar radius, and
obtained $a_{\rm 0}$=0.2. Our model provides a smaller
$a_{\rm 0}=0.006\pm 0.002$ due to the non-zero dipole size
($2a=0.008\pm 0.003$),
that allows
us to take into account the non-symmetrical (with respect to the
stellar center) magnetic field configuration of the star. The
magnetic field intensity and location of the positive and negative
magnetic poles in the stellar rotational reference frame are:

\begin{equation}\label{poles}
\begin{array}{lll}
B_{p}\!=\!3.4\pm 0.7~{\rm kG},&\!\lambda_{p}\!=\!-9^\circ\!\pm 5^\circ,&
\!\delta_{p}\!=\!-33^\circ\!\pm 12^\circ; \\
B_{n}\!=\! -3.3\pm 0.7~{\rm kG},&\!\lambda_{n}\!=\!171^\circ\!\pm 5^\circ,&
\!\delta_{n}\!=\!33^\circ\!\pm 12^\circ.
\end{array}
\end{equation}

\noindent Due to the shift of the dipole from the stellar centre, the positive
magnetic pole has a slightly stronger field intensity than the negative pole,
but not all the analysed lines are equally sensitive to this difference.
Nevertheless, the final model is hardly distinguishable
from a symmetric central dipole model, which provides a
slightly poorer agreement between the simulated and
observed data.
For Fe\,{\sc ii} lines 
the centered dipole
model results in a $\chi^{2}$-function which exceeds by 3\%$\div$5\% the best
fit result from the decentered dipole model. This fact can be related to the quality
of the data and to the model sensitivity.
The dipole offset could be a result of
the particular geometry of 78 Vir,
because we never completely see the positive magnetic pole of the star.
The modeled global magnetic field configuration of 78 Vir is illustrated in
Fig.~\ref{structure} according to the best fit results obtained for Fe\,{\sc ii}
$\lambda$5018\AA.
Five of the Fe\,{\sc ii} lines analysed using the entire Stokes
vector result in a plane-of-sky orientation of the rotational axis of
$\Omega=110\degr\pm17\degr$, while the last two lines (Fe\,{\sc ii}
$\lambda$6432\AA\, and $\lambda$6516\AA), with comparatively weaker variability
of the observed Stokes $Q$ and $U$ profiles, provide higher values of this
angle. Fig.~\ref{XOmega} shows the dependence of the $\chi^{2}_{\rm Q}$ and
$\chi^{2}_{\rm U}$ functions on the angle $\Omega$ for the three
strong Fe\,{\sc ii} lines (see Subsec.~\ref{strong}). Both
functions reach their minima at $\Omega=109\degr$ and $\Omega=289\degr$.
The second value describes a model which is indistinguishable from the configuration
specified in Table~\ref{fe2sm} due to the definition of the Stokes Q and U
parameters.

\begin{figure}[th]
\includegraphics[scale=1.0]{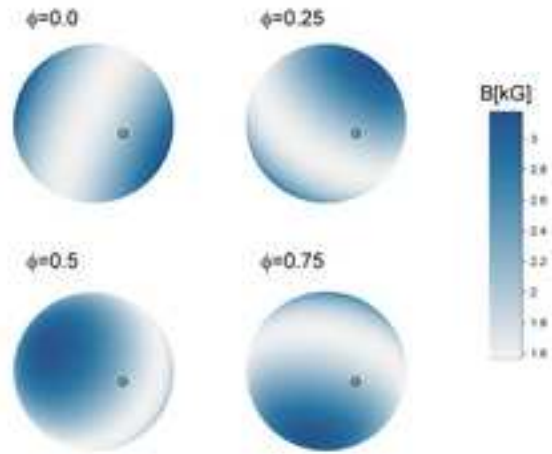}
\caption{Illustration of the surface magnetic field geometry of 78 Vir.
The location of the positive rotation pole is indicated by the open circle.
We always see the negative magnetic pole and only for phase
$\varphi=0.0$ are the positive and negative magnetic poles visible
simultaneously. The stellar image has been rotated in the plane of the
page by the positional angle $\Omega$=110\dr.
\label{structure} }
\end{figure}

As discussed in Sect. 3, 78 Vir is a probable member of the Ursa Major stream.
It approaches us with a mean radial
velocity $V_{\rm r}=-8.1\pm1.0$ km~s$^{-1}$
(Wade~et~al.~2000a). This velocity varies by about $\pm2$ km~s$^{-1}$ with
the phase of stellar rotation (Preston~\cite{Preston69}, this work).
The possible variability of the radial velocity is not taken into account
in the simulation procedure. Respectively, our estimations of the radial
velocity for different Fe\,{\sc ii} lines are distributed around the
aforementioned value within the range $\pm1.5$ km~s$^{-1}$.


\begin{figure*}[th]
\begin{tabular}{@{\hspace{-0.07in}}c@{\hspace{+0.07in}}c}
a) & b) \\
\includegraphics[width=3.5in]{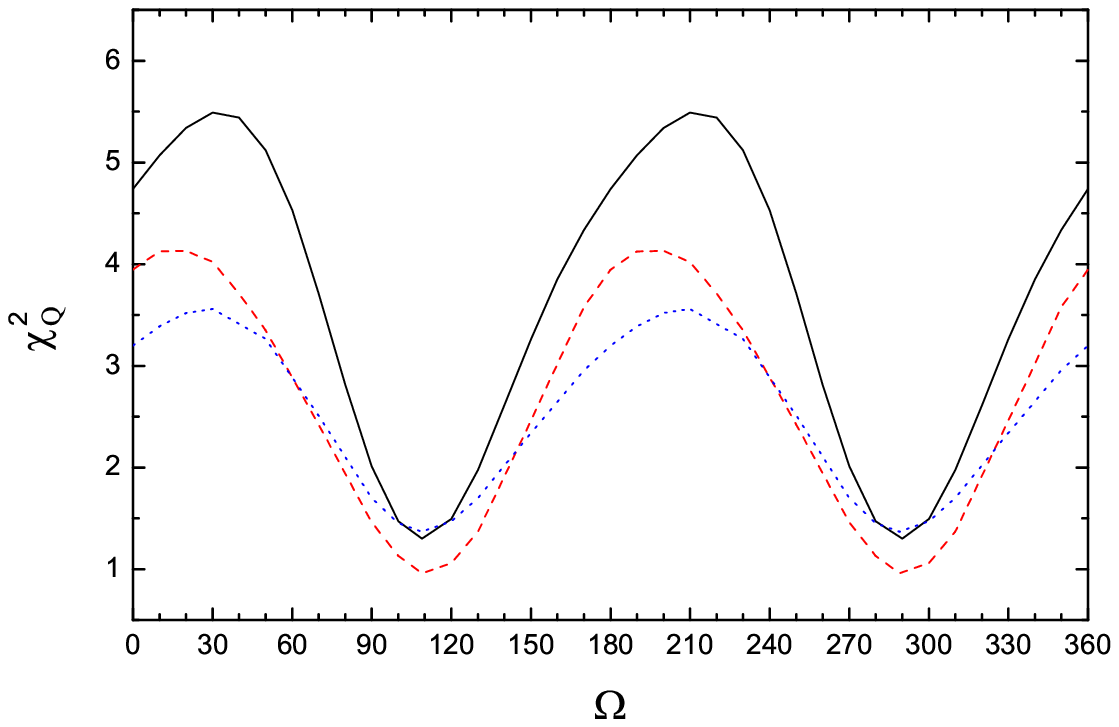} &
\includegraphics[width=3.5in]{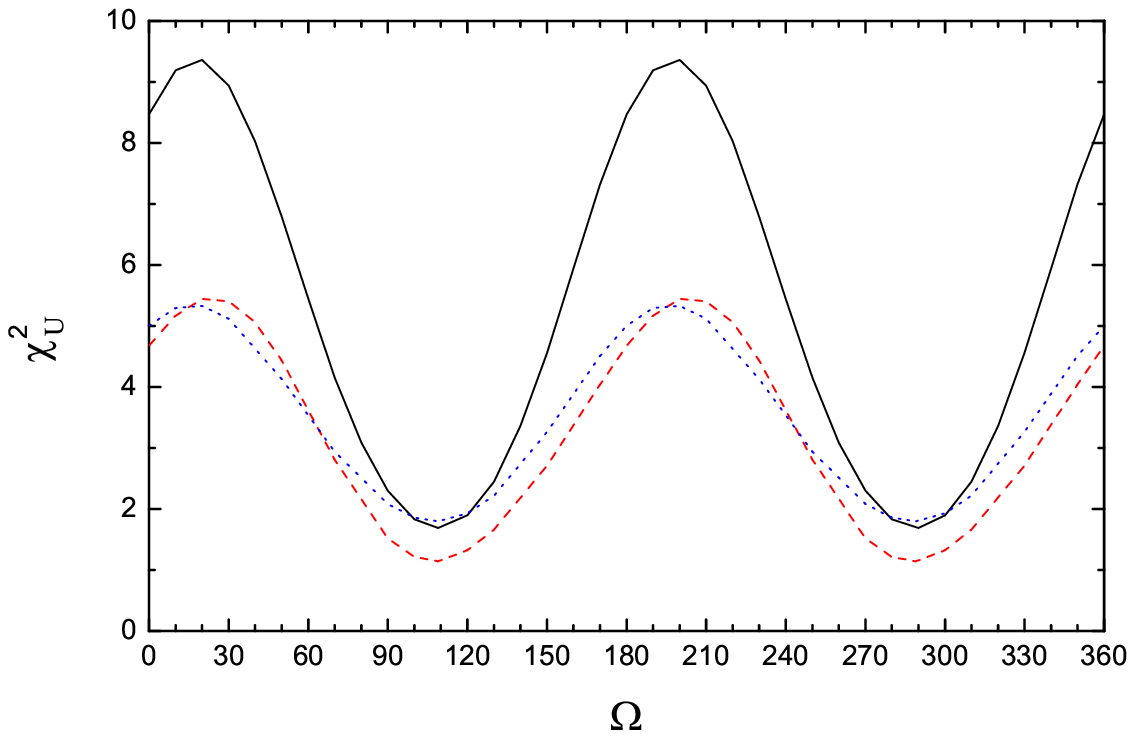}
\end{tabular}
\caption{Dependence of a) $\chi^{2}_{\rm Q}$ and  b) $\chi^{2}_{\rm U}$
on the angle $\Omega$ calculated for three strong lines Fe\,{\sc ii}
$\lambda$4923.927\AA\, (solid line), $\lambda$5018.44\AA\,
(dashed line) and $\lambda$5169.03\AA\, (dotted line). The calculations
were performed under the condition of stability of all the best fit parameters
for these lines (see Table~\ref{fe2sm}) except the $\Omega$. }
\label{XOmega}
\end{figure*}

Nevertheless, the $V_{\rm r}$ variability through the rotational cycle has
been determined from the strong Fe\,{\sc ii} lines $\lambda$4923.927\AA\,
$\lambda$5018.44\AA\, and $\lambda$5169.033\AA\, (see Subsec.~\ref{strong}) and
for the other Fe\,{\sc ii} lines as well. We show that the mean radial
velocity derived from the simulation varies from line to line (see
Table~\ref{fe2sm}). The other Fe\,{\sc ii} lines provide a similar
$\Delta V_{\rm r}$ variability. The precision of the mean radial velocity
determination depends on the observational errors and on the quality of
the description of the surface magnetic field structure. For the blended profiles
it also depends on the accuracy of oscillator strengths.


The analysed lines show weak, coherent variations of
$\Delta V_{\rm r}$ with the period of stellar rotation (see Fig.~\ref{VelocityDiff}).
This suggests a possible moderately non-uniform distribution of Fe\,{\sc ii}.
Nevertheless, the non-uniform iron distribution will not lead to a significantly
different surface magnetic field structure. The simulation of Cr\,{\sc ii} lines
with the assumption of a non-uniform chromium distribution resulted in almost the
same field structure that we obtained from the simulation of Fe\,{\sc ii} lines.


The other best fit parameters are the rotational axis inclination
$i=24\degr\pm5\degr$ and the dipole axis obliquity $\beta=124\degr\pm5\degr$,
which are in a good agreement with the estimates of Leroy~et~al.~(\cite{Leroy+96}).
The derived $V_{\rm e}\sin{i}=12\pm1~{\rm km~s^{-1}}$
provides $V_{\rm e}=29\pm 4~{\rm km~s^{-1}}$ and is consistent with that
of Preston~(\cite{Preston71}).





The derived surface magnetic field variability interval
ranges from 2.1~kG to 3.2~kG and covers the value $B_s$=2.9~kG estimated by
Preston~(\cite{Preston71}). Besides, as Fig.~\ref{Blon1} shows, the longitudinal
magnetic field variation obtained from the two
Fe\,{\sc ii} lines simulation is also in good agreement with the LSD $B_{\rm l}$
data (Wade~et~al.~\cite{Wade+00b}). These facts justify the applicability of the
slightly decentered magnetic dipole model with the aforementioned values of free
parameters for the global magnetic field structure description at 78 Vir.

Unfortunately in the case of 78 Vir the intensity of the Stokes $Q$ and $U$
profiles is similar to the observational errors.
The data are unsuitable for performing an analysis of the small-scale structure of
the magnetic field, using a more sophisticated technique such as
Magnetic Doppler Imaging (MDI; Kochukhov~\&~Piskunov~\cite{K+P02}).
However, it appears, given the relatively good agreement between the intensity,
structure and variability of the observed and simulated Stokes $QU$ profiles,
that the real magnetic field of 78 Vir does not depart strongly from the configuration
derived here. At the same time, we have noted that the Stokes $QU$ profiles are not,
at some phases, fit to within their errors. Therefore these data, notwithstanding their
relatively low S/N, already suggest the limitations of the MCD model framework.


A similar analysis of the Stokes $IV$ profile variability was performed by
Kochukhov et al.~(\cite{Kochukhov+02}) for $\alpha^2$~CVn.
Those authors applied MDI
using ``multipolar regularization'',
analysing the Stokes $IV$ line profiles and obtained a good agreement between
observed and computed profiles for Fe\,{\sc ii},
Cr\,{\sc ii} and Si\,{\sc ii} lines.
The quality of the Stokes $I$ and $V$ profile fits obtained in this
paper is only marginally poorer
than that obtained by Kochukhov et al.~(\cite{Kochukhov+02}), although remarkably
the magnetic field model framework employed here is much simpler.


New spectro-polarimetric observations of 78 Vir with higher spectral resolution and signal-to-noise
ratio (especially for the Stokes $Q$ and $U$ spectra) are required
for further refinement of the global magnetic field
structure. With such data, a model with additional parameters (such as MDI) could be applied,
in order to study the local magnetic field topology and the detailed relationship between
the magnetic field and the abundance distributions.

%
%

\begin{acknowledgements}

The authors are grateful to Prof. John Landstreet and to the referee, Dr. O. Kochukhov,
for their valuable remarks and advice that have led to the improvement of this paper.
The authors acknowledge grant support from the Natural Sciences and Engineering Research Council of
Canada, and the Department of National Defence of Canada (DND-ARP).

\end{acknowledgements}


\end{document}